\documentclass[%
reprint,
nofootinbib,
 amsmath,amssymb,
 aps,
]{revtex4-1}

\usepackage[export]{adjustbox}
\usepackage{color}
\usepackage{graphicx}
\usepackage{dcolumn}
\usepackage{bm}
\usepackage{hyperref}
\usepackage{natbib}
\usepackage{float}
\usepackage[caption=false]{subfig}
\usepackage{hyperref}
\hypersetup{
    colorlinks=true,
    linktoc=page,    
    linkcolor=blue,
    urlcolor=magenta
}

\newcommand{\be}{\begin{equation}}
\newcommand{\ee}{\end{equation}}
\newcommand{\bea}{\begin{eqnarray}}
\newcommand{\eea}{\end{eqnarray}}

\newcommand{\mM}{\mathcal{M}}

\newcommand{\mH}{\mathcal{H}}
\newcommand{\mO}{\mathcal{O}}

\def\/{\frac}
\def\pd{\partial}

\newcommand{\nn}{\nonumber}

\newcommand{\dd}{\,\mathrm{d}}

\newcommand{\ket}[1]{\left|{#1}\right>}

\newcommand{\pmM}{\partial\mathcal{M}}

\newcommand{\LorP}{paper}

\begin{document}
\title{Soft black hole information paradox: Page curve from Maxwell soft hair of a black hole}

\author{Peng Cheng$^1$}
\email{p.cheng@uva.nl}

\author{Yang An$^2$}
\email{anyangpeacefulocean@zju.edu.cn}

\affiliation{
$^1$Institute for Theoretical Physics, University of Amsterdam, 1090 GL Amsterdam, Netherlands\\
$^2$Zhejiang University of Technology (ZJUT), 310014 Hangzhou, Zhejiang, China
}


\begin{abstract}
Treating Maxwell soft hair as a transition function that relates U(1) gauge fields living in the asymptotic region and near-horizon region, the U(1) gauge parameter $\phi(x^a)$ naturally becomes a good label of those Maxwell soft hair degrees of freedom.
This interpretation also builds the connection between Maxwell soft hair and U(1) edge modes living in the intermediate region, which admits a well-defined effective action description.
We study the statistical properties by Euclidean path integral, which concludes that the soft hair density of state increases with black hole temperature.
Hawking radiations increase black hole entropy by creating entanglements, while the measurements of soft modes project the black hole onto lower entropy states.
The competition between phase spaces of Hawking radiations and soft hair measurements gives rise to one version of the Page curve consistent with the unitary evolution of the black hole.
\end{abstract}

\maketitle


\section{Introduction}
\label{intro}

The black hole information paradox (BHIP) was put forward by Hawking in 1976 \cite{Hawking1976}, which ironically seems to suggest Hawking's own model of radiation \cite{HAWKING1974, Hawking1975} should be modified largely at the late time of evaporation.
Even with the recent exciting progress of the  ``Island prescription" \cite{Penington2019, Almheiri2019, Almheiri2019a, Almheiri2019c, Penington2019a, Almheiri2019b, Almheiri2020}, there are no general agreements regarding what has been missed by Hawking's calculation.
Was Hawking missed quantum effects of gravity? Then we might need a while to understand this problem completely.
Or optimistically, if the answer is classical global effects, the problem might be more handleable.

The paradox is that the von Neumann entropy of black holes cannot be larger than the coarse-grained entropy that is proportional to the area of the horizon.
Hawking radiation creates entanglement between the black hole and radiation.
However, in principle, after a certain time of evaporation $t_\text{page}$, no more entanglement can be created, because there is simply no room in the black hole to store the information of the Hawking partner $P$ anymore.
We meet the conundrum that fine-grained entropy becomes larger than coarse-grained entropy if the black hole continues to evaporate after the Page time.
Naively, there are several ways out.
The first one is that the Hawking partner $P$ stored in the black hole can go out of the same horizon as Hawking radiation $H$ to purify the radiation, which is forbidden because of the Almheiri-Marolf-Polchinski-Sully (AMPS) firewall argument \cite{Almheiri2012}.
The second one is that the information of $P$ is encoded somewhere else, which seems not possible because of the no-hair theorem \cite{Israel1967, Israel1968}.
The third one is that the black hole interior is already included in the entanglement wedge of radiation and can be reconstructed from the radiation \cite{Papadodimas2012, Verlinde2012}.
The recently proposed island prescription also belongs to this category.

The recent paper by Pasterski and Verlinde (PV) \cite{Pasterski2020} seems to have unified the above arguments, claiming that the global soft hair beyond the no-hair theorem is the handle to reconstruct interior operators and is firewall-free.
They treated the supertranslation soft hair as a transition function that connects the asymptotic region and near-horizon region, which provided a way to understand the soft hair degrees of freedom.
A dressed infalling observer would perform a measurement of the classical value of the transition function $f(z^A)$ when crossing the intermediate region.
The measurement can project the black hole onto a specific soft hair state that enables the reconstruction of the black hole interior and decreases the black hole entropy.
They also gave seven key assertions that related to the soft hair and checked them in the paper.
However, the supertranslation story did not provide a way to evaluate the phase space of supertranslation soft hair and thus cannot answer if the soft hair is powerful enough to give rise to the Page curve \cite{Page1993} consistent with the unitary evolution of the black hole or not.

Maxwell's theory, on the other hand, has a much simpler structure, and we are able to analyse the phase space quantitatively and get an effective action description of the Maxwell soft hair.
By studying the Maxwell soft hair in the current \LorP, the U(1) gauge transition function is identified with the edge modes living in the intermediate region studied in literature \cite{Donnelly2014, Donnelly2015, Donnelly2016, Blommaert2018, Blommaert2018a}.
The basic idea is the following.
Divide the spacetime into three regions: the near-horizon region $\mM$, the asymptotic region $\bar\mM$, and the intermediate boundary $\pd\mM$; the gauge fields living in $\mM$ and $\bar\mM$ need to match each other on the boundary.
We introduce Lagrange multiplier currents $J^\mu$ along the boundary to match the above gauge fields.
The path integral over $J^{\mu}$ gives out an effective action for $\phi$ which is a massless scalar living on the boundary
\be\label{Sphi}
S[\phi]=-\frac{1}{2 f(L)}\int_{\pmM} d^3x \sqrt{-h}~ \pd^a\phi\pd_a\phi\,.
\ee
Now the would-be gauge parameter $\phi(x^a)$ becomes physical degrees of freedom living on the boundary, and the massless bosonic field can be regarded as the Goldstone modes of the large gauge symmetry on the boundary.

The Hawking radiations increase the entropy of the black hole, and the measurements of the Maxwell soft hair project the black hole onto lower entropy states.
Fermi's golden rule compares the rates of different physical processes by comparing their phase space, which can also be adopted here to compare the Hawking radiation and soft hair measurement.
Equipped with the effective action (\ref{Sphi}), we can directly use the Euclidean path integral to calculate the partition function and evaluate the size of the phase space for Maxwell soft hair.
The competition between the phase spaces of the soft hair and black hole, i.e., the rates of measurements and Hawking radiations, gives out one version of the Page curve consistent with unitarity.

Note that soft hair as a potential solution to BHIP is not new and was already hinted at in many papers, such as \cite{Hawking2015, Hawking2016a, Strominger2017, Strominger2017a, Mirbabayi2016, Bousso2017, Haco2018, Haco2019, Donnay2018, Nomura2018, Nomura2019, Betzios2020, Gaddam2020, Himwich2020}.
The new ingredients here are the Page curve and the analysis of the phase space of soft hair by studying edge modes effective action.

Another motivation of this paper is that people are searching for free parameters \cite{Verlinde2020, Marolf2020, Marolf2020a} that label the superselection sectors inspired by island prescription and the ensemble average proposal \cite{Saad2019, Bousso2020, Penington2019a}.
Swampland program opposes such kind of free parameters for $d>3$ in quantum gravity \cite{Harlow2018, Harlow2018a, McNamara2020, Harlow2020}.
However, approximate global symmetry \cite{Fichet2019} rather than exact ones might survive from swampland and can provide such free parameters.

The remainder of the \LorP is organised as follows.
In section \ref{review}, we provide a brief review of the gravitational story, including properties of supertranslation soft hair, gravitational dressing of operators, and quantum information protocol to reconstruct interior operators.
In section \ref{Esoft}, we present an analysis of the relation between U(1) edge modes and Maxwell soft hair and get an effective action for those Goldstone modes.
The effective action enables us to calculate the statistical properties of those soft hair degrees of freedom.
In section \ref{pagecurve}, we introduce Fermi's golden rule to analyse the rates of two processes that increase and decrease the entropy of the black hole separately.
A version of the Page curve can be gotten from the competition of those two processes.
We end with some conclusions and further comments in the last section.

\section{Review of gravitational story}
\label{review}
Here we provide a brief review of the supertranslation story. The transition function related to the supertranslation between different regions on an asymptotic flat black hole is analysed to make the Maxwell story conceptually easier to accept. We also provide some basic idea of what role soft hair can play in the black hole information paradox by some quantum information protocol.
One can consult \cite{Pasterski2020, Hawking2016a, Strominger2017} for more details.

The origin of the gravitational soft hair is the ambiguity in defining the metric in the asymptotic region. 
For an asymptotically flat Schwarzschild black hole, the asymptotic region is Minkowski spacetime Mink$_4$ as we take $r\to\infty$ limit. 
An isometry is a coordinate transformation that leaves the metric invariant, which is a sign of the spacetime symmetry and corresponds to conserved charges. 
The asymptotic symmetries are the diffeomorphisms that preserve the asymptotic metric because the change of the metric coming from the transformation can also die off as $r\to \infty$. 
It is an extended version of isometry. 
At the linearized level, the asymptotic group is generated by the transformations called supertranslations. 
For a standard Schwarzschild metric in the advanced Bondi coordinates, the diffeomorphisms
\be
g_{\mu\nu}\to g'_{\mu\nu}=g_{\mu\nu}+\mathcal{L}_{\zeta}g_{\mu\nu}
\ee
that preserve Bondi gauge and the standard falloff conditions at the asymptotic region are generated by vector \cite{Hawking2016a}
\be\label{superT}
\zeta_f=f~\pd_v-\frac{1}{2}D^2f~\pd_r+\frac{1}{r}D^A f~\pd_A\,.
\ee
Note that $f$ is a function of the celestial sphere coordinates.

\begin{figure}
  \centering
  \includegraphics[width=6cm]{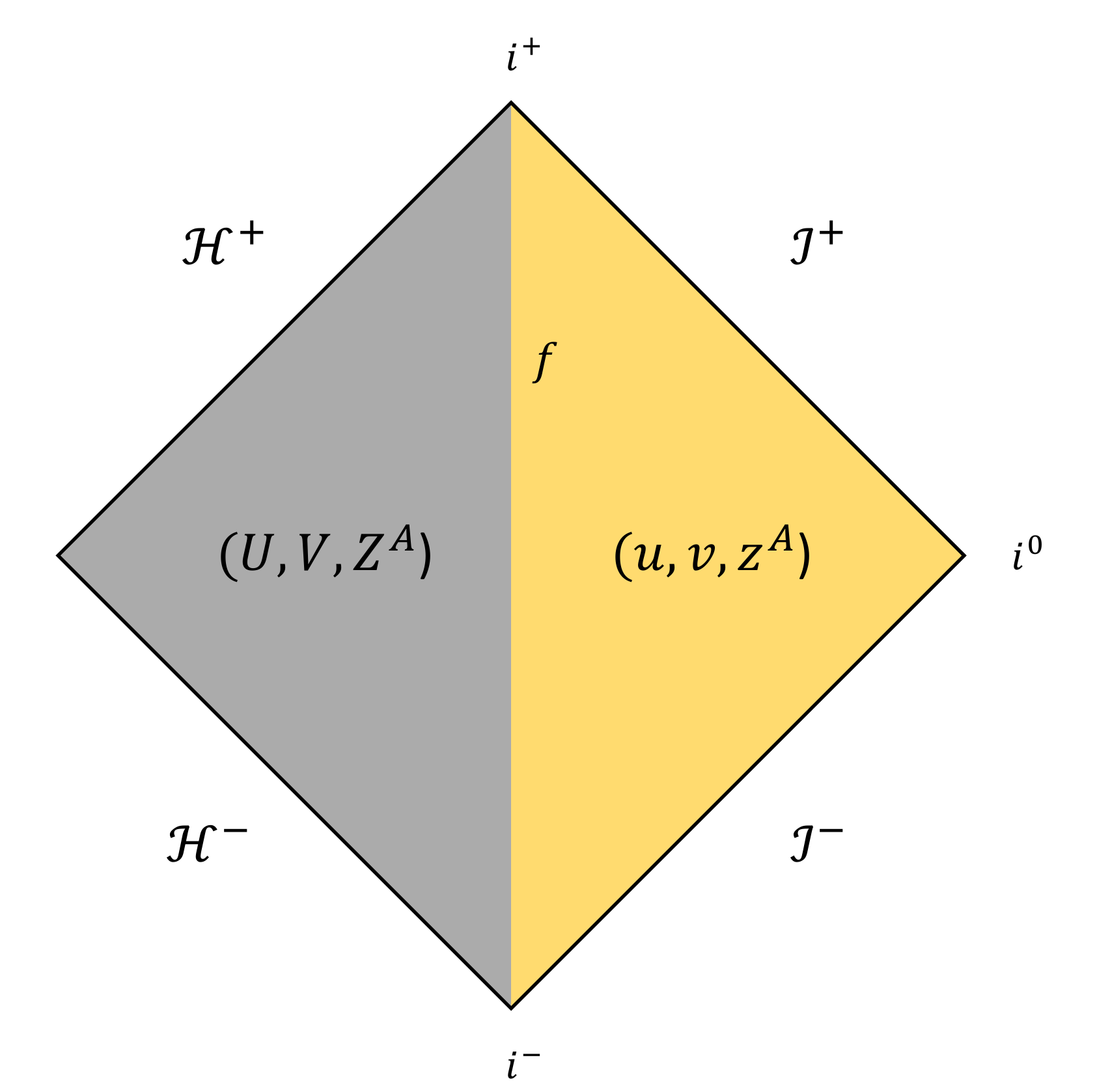}\\
  \caption{Supertranslation soft hair degrees of freedom are encoded in the transition function $f$ that relates different coordinate systems. The relations between those two coordinates are shown in equations (\ref{transf1})-(\ref{transf3}).}\label{matchf}
\end{figure}

\begin{figure}
  \centering
  \includegraphics[width=6cm]{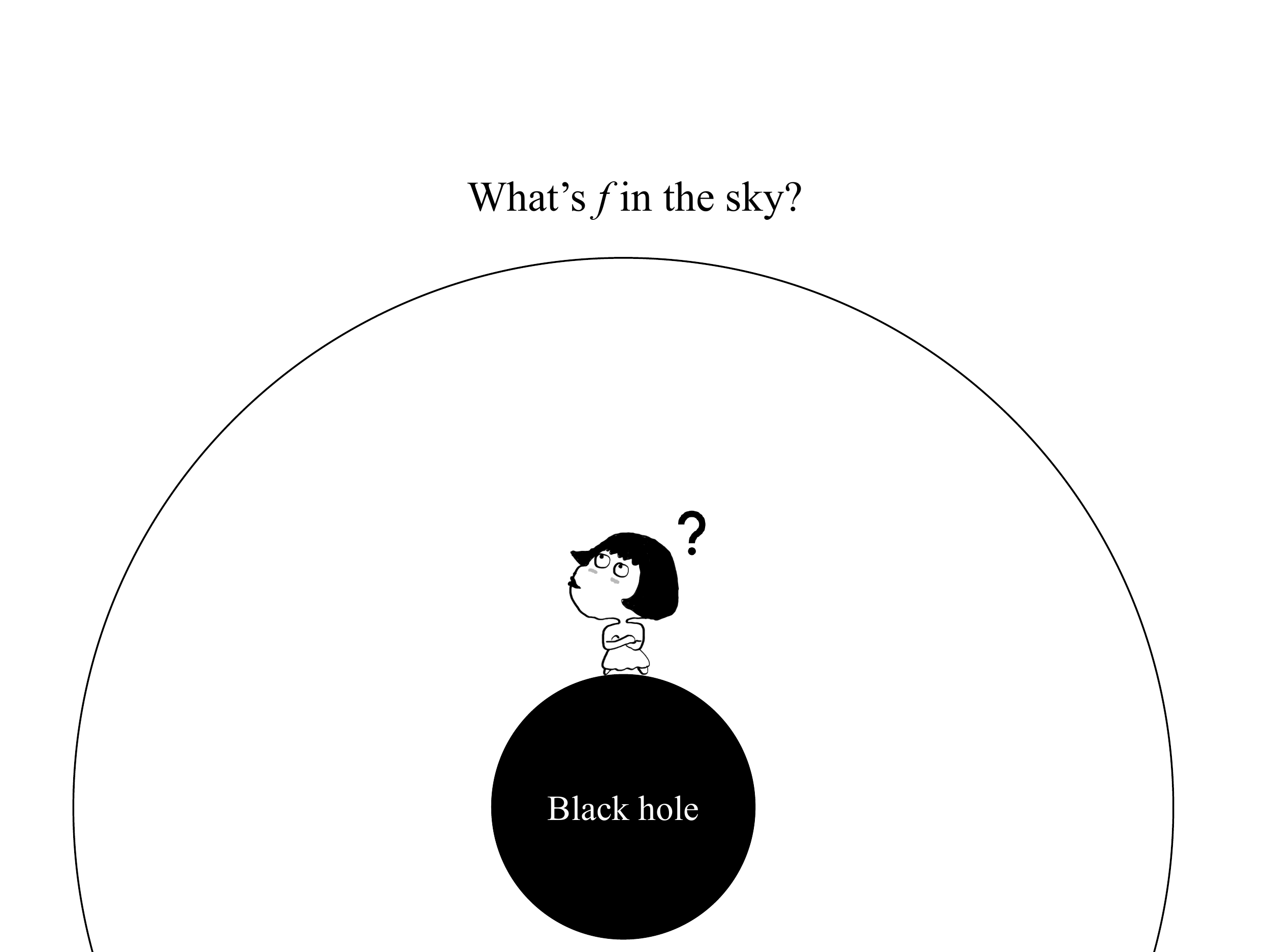}
  \includegraphics[width=6cm]{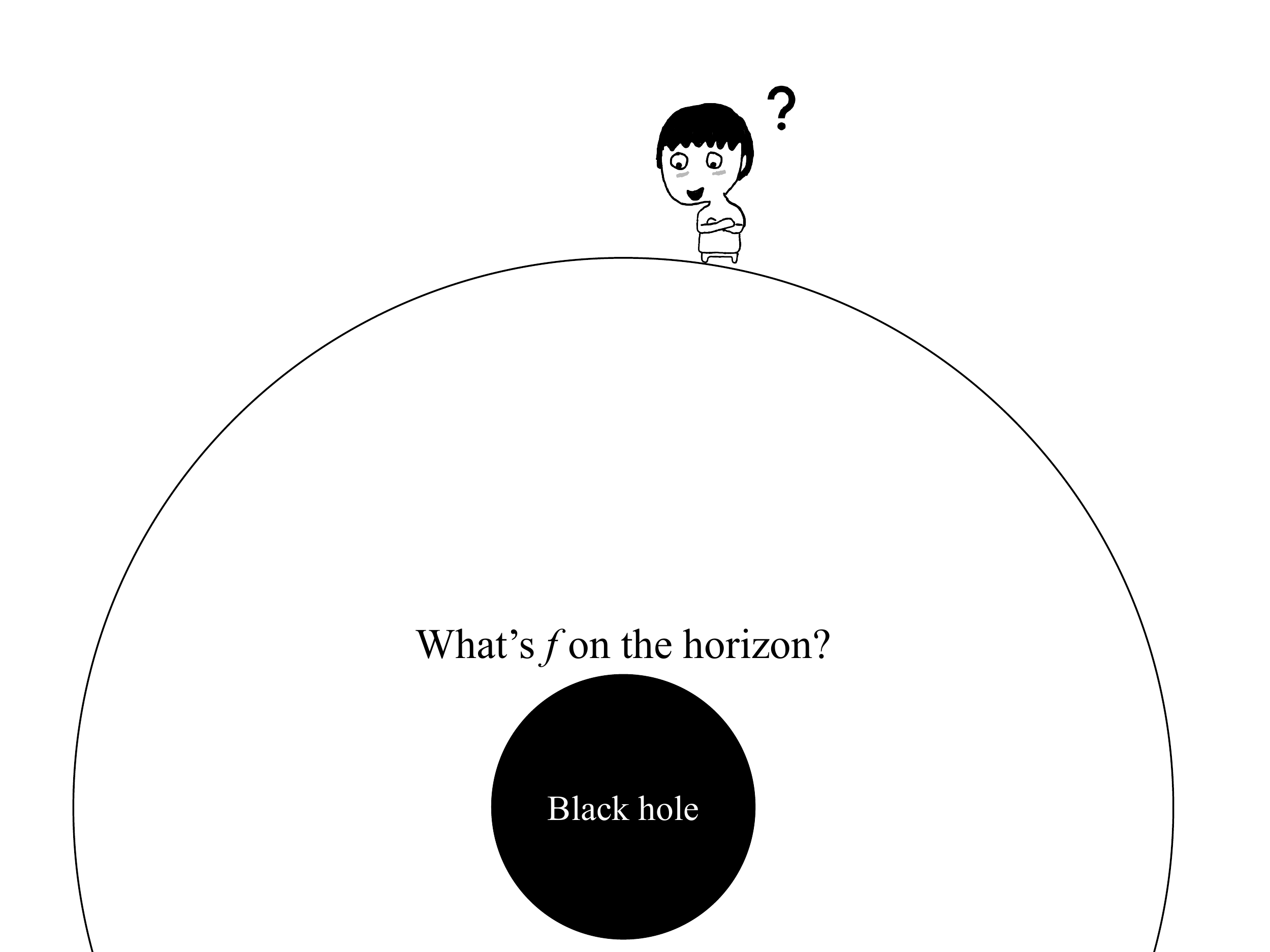}\\
  \caption{Schematic depiction of asymptotic symmetry group and soft hair. \textbf{First panel}: for an observer sitting on the horizon, there is an ambiguity in the sky generated by the diffeomorphisms that preserve the asymptotic structure. This ambiguity is labelled by a parameter $f(z^A)$ and is a global effect.
\textbf{Second panel}: an asymptotic observer meets a similar problem. The black hole is a superposition state of different $f(z^A)$s unless the observer jumps into the black hole to measure it. By jumping across the boundary, the black hole is projected to a state with more specific $f$, which has lower entropy.}\label{softf}
\end{figure}

One of the crucial insights of \cite{Pasterski2020} is that one can change perspective and characterise that diffeomorphism in terms of a transition function that connects the asymptotic and near-horizon coordinates. 
We choose light cone coordinates $(u,v)$ in the asymptotic region such that the metric is
\be\label{asymp}
ds^2\big{|}_{\text{asymp}}=\Lambda du dv+r^2\gamma_{AB}dz^A  dz^B\,,~\Lambda \equiv 1-\frac{2M}{r}\,,
\ee
with $\gamma_{AB}$ the metric on $S^2$. 
In the near-horizon region, we use Kruskal-Szekeres coordinates $(U,V)$ and write the metric as
\be\label{hor}
ds^2\big{|}_{\text{hor}}=-FdUdV+r^2\gamma_{AB}dZ^AdZ^B\,,~ F\equiv \frac{2M}{r}e^{-r/2M}\,.
\ee
Those two coordinates can be patched together via coordinate transformation
\be\label{coord}
\frac{U}{4M}=-e^{-u/4M}\,,~~
\frac{V}{4M}=-e^{v/4M}\,,~~
Z^A=z^A\,.
\ee
There is ambiguity in matching those two coordinates, and thus the transformation can be modified by adding linearized soft hair into
\bea
\frac{U}{4M} &=& -e^{-u/4M}+\zeta_f^U\,,\label{transf1}\\
\frac{V}{4M} &=& -e^{v/4M}+\zeta_f^V\,,\label{transf2}\\
Z^A &=& z^A+\zeta_f^A\,.\label{transf3}
\eea
Here $\zeta_f$ is exactly the components of the vector field associated with the supertranslation in equation (\ref{superT}). 
The above coordinate transition between the two coordinate systems (\ref{asymp}) and (\ref{hor}) can also be written as \cite{Pasterski2020}
\bea\label{trans}
v&=& 4M\ln\frac{V}{4M}-f\,,\\
u &=& -4M\ln(-\frac{U}{4M})-f-\frac{1}{F}e^{(u-v)/4M}D^2f\,,\\
z^A &=& Z^A-\frac{1}{r}D^Af\,,
\eea
where $D_A$ is the covariant derivative with respect to $\gamma_{AB}$.
The transition function $f$ that relates different coordinate systems is schematically depicted in figure \ref{matchf}. Now, as shown in figure \ref{softf}, there are extra degrees of freedom in defining the asymptotic structure for the observers sitting on the horizon, which are labelled by the supertranslation parameter $f(z^A)$. 
Similarly, the asymptotic observers also have problems in deciding what soft hair state the black hole is. Those extra degrees of freedom are the gravitational soft hair. 

The gravitational soft hair discussed above is physical degrees of freedom rather than redundancy in the description. 
There are conserved charges associated with the supertranslation and canonical conjugate momentum dual to the parameter $f(z^A)$ through standard symplectic form analysis \cite{Strominger2017}. 
Promoting $f$ to a quantum operator, $\hat f(z^A)$ describes the soft hair Goldstone modes. 
The soft charge $Q_S(z^A)$ can be written as
\be
Q_S(z^A)=\int dv~\hat q_S(v,Z^A)\,,
\ee
and conjugate variable to $\hat f$ can be seen from the commutator
\be
[\hat q_S(v,z^A),f(v',z'^A)]=i\delta(v-v')\delta^{(2)}(z-z')\,,
\ee
with
\be
\hat q_S(v,z^A)=\frac{1}{16\pi}D^2(D^2+2)~\pd_v \hat f\,.
\ee
Also, one can implant soft hair by sending in shock waves at $v=v_0$ with some specific stress tensors \cite{Hawking2016a}
\bea
\hat T_{vv} &=& \frac{1}{16\pi M^2}\left[m-\frac14 D^2(D^2-1)\hat f\right]~\delta(v-v_0)\,,\\
\hat T_{vA} &=& \frac{3}{32\pi M}D_A\hat f~\delta(v-v_0)\,.
\eea
where $m$ is the mass of the shock wave.
The gravitational soft hair can be changed by the shock wave stress tensor
\bea
g_{\mu\nu}\to g'_{\mu\nu}=g_{\mu\nu}+\Theta(v-v_0)\left( \mathcal{L}_{f}g_{\mu\nu}+\frac{m}{r}\delta^v_a\delta^v_b \right)\,,
\eea
which means that $\hat f$ are changed to $\hat f+f$ by the shock wave.
Implanting supertranslation soft hair is another physical argument showing that soft hair is not a description redundancy.

In a gravitational theory, diffeomorphism invariant (physical) operators must commute with the total supertranslation charge $Q_f=Q_S+Q_H$
\be
[Q_f,\mO_{phys}]=0\,,
\ee 
which requires the physical operators factorise into a product of matter operator $\mO$ times a gravitational Wilson line $\mathcal{W}$, i.e., 
\be
\mO_{phys}=\mO \times \mathcal{W}\,.
\ee
The Wilson line $\mathcal{W}$ takes the following form \cite{Himwich2020}:
\be
\mathcal{W}(k,z^A)=e^{-ik\hat f(z^A)}\,.
\ee
For an infalling operator $\mO(v,z^A)$, the gravitational dressing is just adding an extra phase on the momentum eigenstate $e^{ikv}$. So the dressed infalling operator can be simply expressed by replacing the original $v$ coordinate by 
\be
\hat v=v-\hat f\,.
\ee
Under the action of $Q_S$, $\hat f$ is shifted by an amount $f$,
\be
\hat f\to \hat f+f\,.
\ee
And the $v$ coordinate of a black hole with gravitational soft hair implanted is shifted with the opposite amount $-f$,
\be
v\to v-f\,
\ee
 as shown in (\ref{trans}). Those two shifted phases cancel each other, and we can conclude that the dressed infalling external operators and the black hole soft hair know the phases of each other. In this sense, the outside observers and black hole soft hair are entangled with each other, which is essential for the reconstruction of interior operators.

\begin{figure}
  \centering
  \includegraphics[width=7cm]{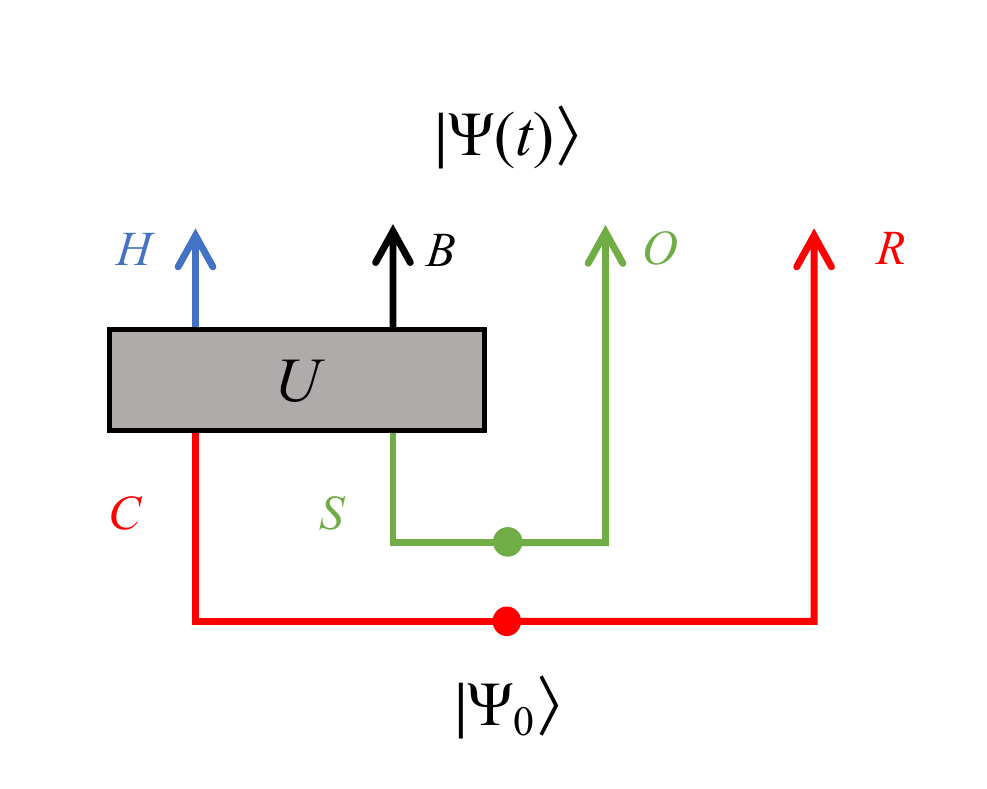}\\
  \includegraphics[width=7cm]{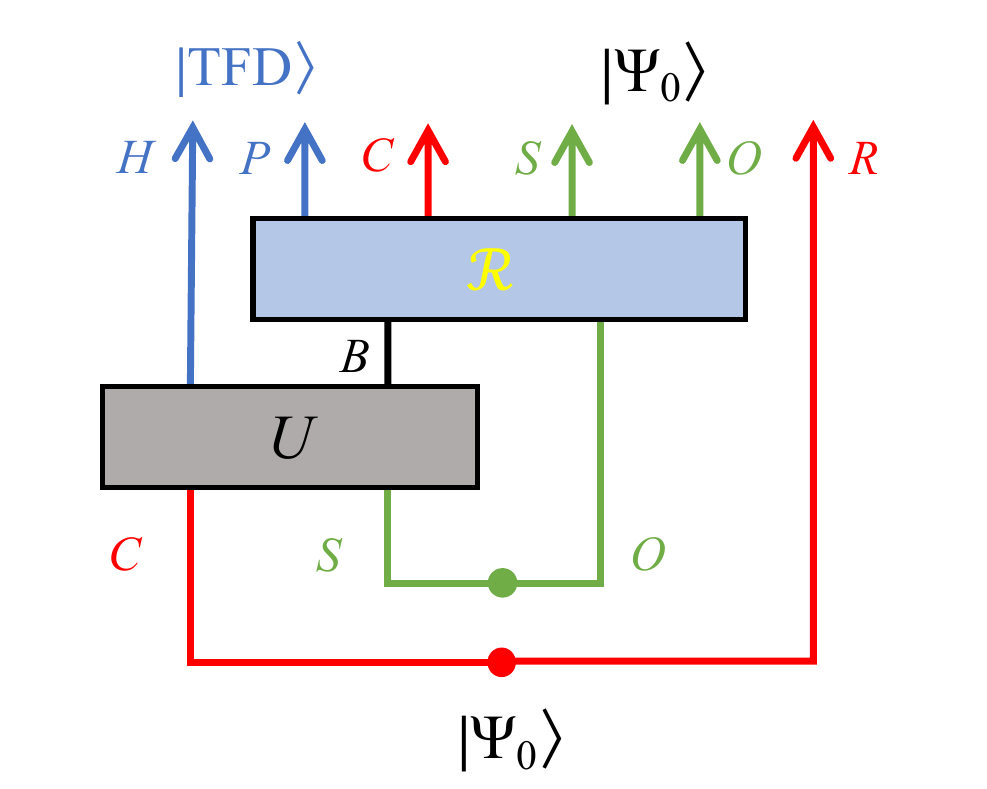}\\
  \caption{Diagrammatic representation of the black hole evolution. \textbf{First panel}: a system including code subspace $C$, early radiation $R$, soft hair $S$, and observer $O$ evolve for a period of Hawking evaporation. \textbf{Second panel}: including the information of observer Hilbert space, the recovery operator $\mathcal{R}$ reverse the evolution $U$, and reconstructed Hawking partner $P$.}\label{U}
\end{figure}

Now with the soft hair degrees of freedom included in the system, we can reconstruct the Hawking partner of the late time radiation using those soft hair degrees of freedom as far as the Hilbert space of soft hair is large. Let us first clarify the notations, and then we can explain the basic idea. We denote the black hole Hilbert space as $\mH_B$ with its size $d_B=\dim \mH_B$, code subspace $\mH_C$ with $d_C=\dim \mH_C$, Hilbert space associated with the observers $\mH_O$ with $d_O=\dim \mH_O$, and similar pattern for early radiation $R$, late time radiation $H$, and mirror degrees of freedom $P$ playing the role of the Hawking partner. 
As discussed in the previous paragraph, $S$ and $O$ purify each other. 
The basis states of code subspace $C$ and radiation $R$ can be represented as $\ket{i}_C$ and $\ket{i}_R$, the ones for soft hair $S$ and observer $O$ are $\ket{f}_S$ and $\ket{f}_O$, and the ones for $H$ and $P$ are $\ket{n}_H$ and $\ket{m}_P$. 

The state we are starting with is
\be
\ket{\Psi_0}=\frac{1}{\sqrt{d_C d_S}}~\sum_{i,f} \ket{i}_C\ket{f}_S\ket{f}_O\ket{i}_R
\ee
which is diagrammatically shown in figure \ref{U}. After embedding the state $\ket{\Psi}_0$ into black hole Hilbert space and time evolving the black hole system to emit one late time Hawking radiation $H$, the wave function of the system can be written as
\be
\ket{\Psi(t)}=\sum_{i,f,n}~C_n T_f\ket{i}_C \otimes \ket{n}_H\ket{i}_R \ket{f}_O\,,
\ee
where $T_f$ is the embedding tensor mapping the code subspace to the Hilbert space $\mH_f$ with a fixed soft hair eigenvalue $f$, and $C_n$ is the Kraus operator. 
The evolution is illustrated in the first panel of figure \ref{U}.

Inspired by the quantum error correction protocol and Petz map, the recovery map $\mathcal{R}$, which reverses the unitary evolution $U$ and reconstructs the code subspace, and the Hawking partner can be worked out. The role of the recovery map is illustrated in the second panel of figure \ref{U}, which can be represented as
\be
\mathcal{R}~U\ket{\Psi_0}\simeq \ket{\Psi_0}\ket{\text{TFD}}_{HP}\,,
\ee
where $\ket{\text{TFD}}_{HP}$ is the thermofield double (TFD) made by Hawking radiation and its partner.
The error of the reconstruct ${d_cd_H}/{d_B}\ll1$ is indicated by the sign $\simeq$. The recovery map within a fixed soft hair sector $\mathcal{R}_{f}$ can be expressed as \cite{Pasterski2020}
\be
\mathcal{R}_f\ket{\Phi}_B=\sum_{f,n}\mathcal{R}_{f,n}\otimes\ket{\Phi}_B \ket{f}_S\ket{n}_P
\ee
where $\ket{\Phi}_B$ is the black hole state.
$\mathcal{R}_{f,n}$ can be expressed in terms of the density matrix of black hole and observer in a fixed soft hair sector $\sigma_{Bf}$
\be
\mathcal{R}_{f,n}=\frac{1}{\sqrt{d_C}}T^\dagger_f C^\dagger_n (\sigma_{Bf})^{-1/2}\,,~~~~~~\sigma_{Bf}=\frac{1}{d_C}\sum_n C_n T_f T_f^\dagger C_n^\dagger\,.
\ee
It can be shown that 
\be
\mathcal{R}_{f,n}C_m T_f\ket{i}_C\simeq \sqrt{p_n}\delta_{nm}\ket{i}_C\ket{f}_S\,,
\ee
where $p_n$ are the Boltzmann weights.
Then we can accomplish our task by the following procedures
\bea
\mathcal{R}~ U\ket{\Psi_0} =&& \frac{1}{\sqrt{d_C d_S}}\,\nn\\ &&\times \sum_{n,m,i,f}\mathcal{R}_{f,n}C_m T_f\ket{i}_C \otimes \ket{f}_O\ket{i}_R \ket{n}_P \ket{m}_H\,\nn\\
\simeq && \frac{1}{\sqrt{d_C d_S}}\,\nn\\ &&\times\sum_{i,f}\ket{i}_C\ket{f}_S\ket{f}_O\ket{i}_R\otimes \sum_n\sqrt{p_n}\ket{n}_H\ket{n}_P\,\nn\\
=&& \ket{\Psi_0}\otimes\ket{\text{TFD}}_{HP}\,,
\eea
as shown in figure \ref{U}.

The fact that the above procedure can be done without AMPS firewall can be reasoned by Yoshida's decoupling theorem \cite{Yoshida2019}, which can be expressed as follows: if $U$ is scrambling and the dimension of $\mH_S$ is much larger than $\mH_H$, i.e., $d_S\gg d_H$, early and late Hawking radiation are decoupled, and one can reconstruct the Hawking partner $P$ without using the early radiation $R$. The Hawking-Perry-Strominger soft hair introduces an observer-dependent firewall \cite{Pasterski2020}, and the infalling observer will never knock into any AMPS firewall before reaching the singularity. 

Several key assertions \cite{Pasterski2020} related to the soft hair degrees of freedom are on the horizon with all the above arguments.
We summarise those assertions below.
The soft hair degrees of freedom are encoded in the transition function (or diffeomorphism) $f(z^A)$; thus, they are classical and measurable properties of black holes, as can be seen from figure \ref{matchf}.
The information carried by the function $f$ should be regarded as part of black hole entropy. 
The observers who are sitting at the horizon or asymptotic region cannot determine what $f$ is; thus as shown in figure \ref{softf} the soft hair is invisible to them. 
Only by jumping across the intermediate region and comparing the coordinate on both sides, the dressed infalling observer adopts the so-called sharp focus perspective. As a result, the black hole is projected to a state with a more specific value of $f(z_0^A)$ on location $z_0^A$. 
The measurement of soft hair would project the black hole onto a soft hair eigenstate with less entropy and reduce the total entropy of the black hole. 
The measurement enables the reconstruction of the Hawking partner using a set of quantum information protocols as discussed above.
Now the entropy of the black hole is lower than before because of this measurement. 
However, the soft hair is not necessarily completely projected out by one measurement because $f(z^A)$ can be a lot of configurations rather than a global parameter. 
So we can gradually reduce the total entropy by repeatedly throwing gravitationally dressed operators across the boundary. 

Now we have encoded the gravitational soft hair in terms of transition function shown in equations (\ref{transf1})-(\ref{transf3}), which makes the definition of the Maxwell soft hair very transparent. 
Although one can largely mimic the gravitational falloff analysis by imposing some falloff boundary conditions on gauge fields near the infinity and claiming that all the gauge transformations consistent with the falloff boundary conditions can be regarded as physical symmetries. 
Here we just use the gauge field to replace the metric and use gauge transformation to replace diffeomorphism. 
However, there is a conceptual difference between the isometry that preserves some metric structure (even asymptotic metric) and other diffeomorphisms. 
This difference is not obvious in gauge theories.
Gauge parameters can have arbitrary dependence on spacetime coordinates $x^\mu$, unless the canonical analysis tells us which of them are physical and which are not.
Understanding the whole story by looking at the transition function and getting an effective action for those would-be gauge degrees of freedom seems more straightforward. Moreover, the transition function interpretation has a great potential to be generalised to a finite distance away from the horizon.
So the strategy for Maxwell's theory is to regard the Maxwell soft hair as the transition function that compares gauge fields living in different regions.
We will explain more details in the next section.

\section{Maxwell soft hair}
\label{Esoft}

In this section, we adopt the language reviewed in the previous section, where we treated gravitational soft hair as a transition function between the asymptotic region and near-horizon region of the global black hole spacetime.
Whereas here, we mainly focus on U(1) gauge theory living on a black hole background and characterise the Maxwell soft hair in terms of a transition function of gauge fields between those different regions.
The main difference is that the show's leading role is the U(1) gauge transformation $\phi(x^a)$ on the boundary $x^a$, rather than supertranslation.
The virtue of Maxwell's theory, in addition to being easier to be handled both conceptually and computationally, is that we can relate it with U(1) edge modes and have an effective Lagrangian description of $\phi$, which can help us to estimate the size of phase space of Maxwell soft hair.
Once we know the size of phase space, we are enabled to do more analysis on what soft hair can do to help us understand BHIP as an example, which will be the central subject of the next section.

\subsection{Maxwell soft hair as a transition function}

\begin{figure}
  \centering
  \includegraphics[width=6cm]{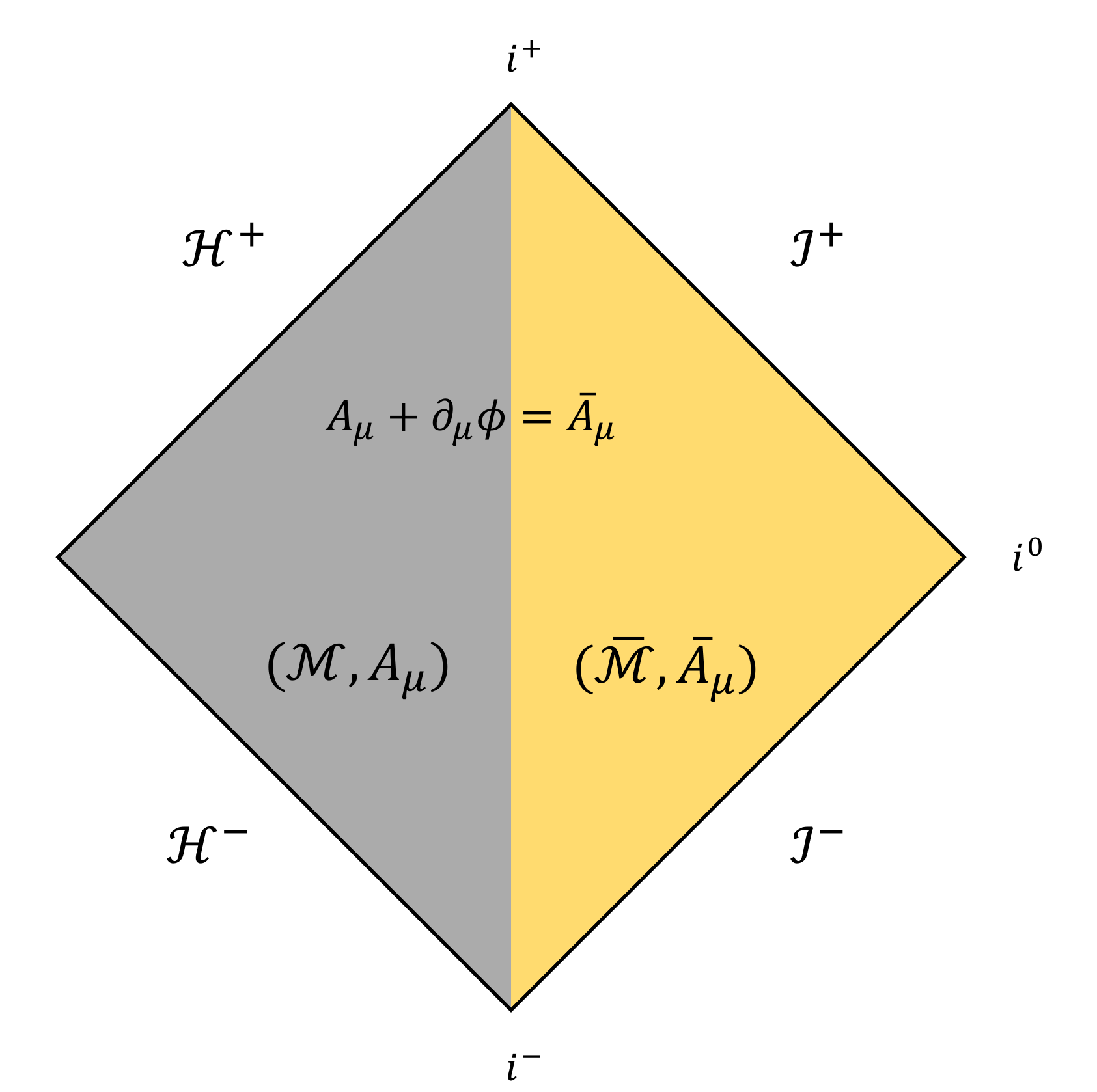}\\
  \caption{Maxwell soft hair degrees of freedom are encoded in the gauge transformation parameter $\phi$ between the gauge fields living in two different regions.}\label{matchphi}
\end{figure}

Similar to the gravity story, we divide the asymptotically flat Schwarzschild black hole into two regions, namely the near-horizon region $\mM$ and the asymptotic region $\bar{\mM}$.
At the place where those two regions meet each other, we interpret the Maxwell soft hair as the transition function that uniformises the gauge fields $A_\mu$ and $\bar A_{\mu}$ living in $\mM$ and $\bar{\mM}$ respectively, as shown in figure \ref{matchphi}.
In this sense, one can think there is a boundary, denoted as $\pd\mM$, between the two regions, and the Maxwell soft hair can be regarded as the edge states living on the boundary \cite{Blommaert2018, Blommaert2018a}.
One can imagine this boundary as some kind of ``Dyson sphere" outside the black hole, and we will mainly adopt the perspective of the observer living inside the Dyson sphere and regard the soft hair degrees of freedom as vacuum degeneracy in the sky.
This perspective is shown in the first panel of figure \ref{softf}; as the observer looks up into the sky, the U(1) soft hair is labelled by gauge parameter $\phi$ on the boundary.

Before taking any further step, let us first set up our physical background.
We are mainly interested in the U(1) gauge theory with action
\be
S=-\frac{1}{4}\int_{\mM}d^4x\sqrt{-g}~F^{\mu\nu}F_{\mu\nu}\,.
\ee
The theory is put on an asymptotically flat black hole spacetime with metric
\be
ds^2=-(1-\frac{2GM}{r})dt^2+(1-\frac{2GM}{r})^{-1}dr^2+r^2d\Omega^2_2\,.
\ee
The spacetime is divided into near-horizon region $\mM$ and asymptotic region $\bar{\mM}$. The path integral on this manifold can be written as
\be\label{pintegral}
\int[DA_\mu][D\bar A_{\mu}]\prod_{x\in \pd\mM}\delta(A_\mu-\bar A_\mu)~e^{i(S_A+S_{\bar A})}\,.
\ee
The gauge fields living on the sides are supposed to match each other. That is the reason why we introduce a boundary delta function in the path integral.
The boundary delta function can be dealt by introducing Lagrange multiplier fields on the boundary
\be\label{LM}
\prod_{x\in \pd\mM}\delta(A_\mu-\bar A_\mu)=\int[DJ^{\mu}]e^{i\int_{\pmM}J^{\mu}(A_\mu-\bar{A}_{\mu})}\,.
\ee
Lagrange multipliers $J^{\mu}$ can be arbitrary functions on the boundary, which will be integrated over eventually.
This part can be regarded as the duplication of coordinate transformation in equation (\ref{coord}).
For the observer living in the near-horizon region, the original gauge theory can be gotten by getting rid of the contributions from $\bar A_{\mu}$.
$A_{\mu}$ should be fully gauge fixed in bulk $\mM$ for a physical observer, and we can write the gauge field as $A_{\mu}-i\pd_\mu \phi$ on the boundary to reserve the ambiguity to match the fields between two sides; hence $\phi$ should be interpreted as the transition function that connects the gauge fields living in those two regions.
This gauge parameter $\phi$ along the boundary has also been interpreted as U(1) edge modes \cite{Blommaert2018, Blommaert2018a}. One can also perform symplectic form analysis to show those modes $\phi$ are indeed physical \cite{Strominger2017}.

Following the same logic as the supertranslation case, we interpreted the gauge parameter as the transition function that relates the Maxwell gauge fields between two regions.
Thus we can say that $\phi$ at the boundary provides a good label of Maxwell soft hair just as supertranslation parameter $f$ did for gravitational soft hair.
Moreover, we also provide a concrete relation between the Maxwell soft hair and the would-be gauge edge states living on the boundary.
This relation was also mentioned in \cite{Blommaert2018}.

The Maxwell soft hair should have the same seven assertions as supertranslation soft hair discussed by PV.
Those properties are essential for soft hair to shed light on the black hole information paradox.
The Maxwell soft hair degrees of freedom should also be classical and physical effects of the global black hole, which are invisible for observers restricted in the asymptotic region or near-horizon region.
Maxwell soft hair can be measured and is exponentially sensitive for an infalling observer who crosses the intermediate region.
Those soft degrees of freedom carry a large amount of entropy and can be projected to a lower entropy state by measurements done by the dressed in-falling particles.
We will use those assertions for Maxwell's theory as the key ingredients to argue the Page curve in the next section.

\subsection{Effective field theory of Maxwell soft hair}

Now we have interpreted the Maxwell soft hair as a transition function that relates the U(1) gauge fields living in different regions.
Moreover, we have built the connection of Maxwell soft hair and the U(1) edge modes living on the intermediate boundary.
The virtue of Maxwell soft hair is that we have an effective action description of those edge modes.
With the help of the action, we can do some quantitative analysis of the phase space of Maxwell soft hair $\phi(x^a)$.

Restricted to the near-horizon region $\mM$, the effective action for $A_{\mu}$ can be written as
\bea
S_A &=& -\frac{1}{4}\int_\mM d^4x \sqrt{-g}~ F^{\mu\nu}F_{\mu\nu}\nn\\
&~& +\int_{\pmM}d^3x\sqrt{-h}~ J^\mu(A_\mu-i\pd_\mu\phi)\,,
\eea
Here $J^\mu$ are the Lagrange multipliers introduced in equation (\ref{LM}) and will be integrated over later.
In order to separate the effects from edge modes and bulk modes, we are going to separate $A_\mu$ into two parts
\be
A_{\mu}=\tilde{A}_{\mu}+B_{\mu}\,,
\ee
where $\tilde{A}_{\mu}$ vanishes at boundary, and $B_{\mu}$ is on shell in the bulk and takes the same boundary value as $A_{\mu}$.
Then the effective action can be separated into two parts
\bea
S_{A} &=& S_{\tilde A}+S_{B}\,\\
&=& -\frac{1}{4}\int_\mM d^4x \tilde F^{\mu\nu} \tilde F_{\mu\nu} \nn\\
&~& +\int_{\pmM} d^3x \sqrt{-h} \left[-\frac{1}{2}n_{\nu} F_{(B)}^{\mu\nu}B_{\mu} + J^\mu(B_\mu-i\pd_\mu\phi)\right]\,,\nn
\eea
where $n^{\mu}$ is the normal vector orthogonal to the boundary and $F_{(B)}^{\mu\nu}$ is the field strength calculated from $B_\mu$.
The first part $S_{\tilde A}$ that captures the bulk fluctuation contribution is not of interest here.
The second part
\be\label{SB}
S_{B}=\int_{\pmM} d^3x \sqrt{-h} \left[-\frac{1}{2}n_{\nu} F_{(B)}^{\mu\nu}B_{\mu} + J^\mu(B_\mu-i\pd_\mu\phi)\right]
\ee
captures the interesting physical effects and will give out the effective action for Goldstone modes $\phi$.
$B_{\mu}$ is on shell in the bulk $\mM$, and thus we can solve $B_{\mu}$ in terms of boundary current $J^{\mu}$.
Putting back the solution $B_{\mu}[J^\mu]$ means that the effective action $S_{B}$ can be written as a functional of $J^{\mu}$ and $\phi$.
To do that, one needs to solve the following bulk problem
\bea\label{problem}
\nabla_\mu F_{(B)}^{\mu\nu}=0\,,~~~~~\text{with}~~~~~n_{\nu} F_{(B)}^{\mu\nu}\big{|}_{\text{bdy}}=J^\mu\,.
\eea
Until now, we have not picked any specific gauge fixing condition.
No matter what gauge fixing condition we pick for $B_{\mu}$ in bulk, $\tilde A_{\mu}$ should use the same gauge fixing condition.
Here we are going to let $B_r=0$ (i.e. $A_r=0$) as our bulk gauge fixing condition.
Choosing $r=L$ as our the boundary, the boundary condition in equation (\ref{problem}) is
\be
\pd_r B_a \big{|}_{\text{bdy}}=J_a\,,~~~~~0=J_r\,,
\ee
where $x^a$ is the coordinates along the boundary.
Working with our gauge fixing condition, one can immediately see that after variable separation, $B_a$ always take the form of
\be\label{solution}
B_a=f(r)\cdot J_a\,.
\ee
$f(r)$ can be some very complicated function of $r$ determined by the bulk equation of motion, with $\pd_r f(r)\big{|}_{r=L}=1$.
This is already enough information for us to get the effective action for Goldstone $\phi(x^a)$. Putting (\ref{solution}) back into the action (\ref{SB}), we get
\be
S_B=\int_{\pmM} d^3x \sqrt{-h} \left(\frac{f(L)}{2}J^a J_a-iJ^a\pd_a\phi(x^a)\right)\,.
\ee
Functional integrating out $J^a(x^b)$ in the path integral, one gets an effective action for $\phi(x^a)$, which is read as
\be\label{action}
S[\phi]=-\frac{1}{2 f(L)}\int_{\pmM} d^3x \sqrt{-h}~ \pd^a\phi\pd_a\phi\,.
\ee
The action describes a three-dimensional massless scalar field living on the boundary $r=L$ with a coupling constant that varies with the location of the boundary.
This is the effective action description of Goldstone modes $\phi(x^a)$.

Now we are good to analyse how large the phase space of Maxwell soft hair is\footnote{A more careful analysis of those Goldstone modes will be treated in a separate paper \cite{Cheng2021}}.
We analyse the statistical properties of $\phi(x^a)$ by performing a Euclidean path integral at finite temperature.
The partition function for $\phi$ can be written as
\be
Z_\phi=\int[D\phi]~e^{-S_E[\phi]}
\ee
where the Euclidean action is
\be\label{Eaction}
S_E[\phi]= \frac{1}{2 f(L)}\int_0^\beta d\tau \int d^2x \sqrt{h}~ \pd^a\phi\pd_a\phi\,,
\ee
with inverse temperature $\beta$ as the periodicity of $\tau$ direction. This is nothing strange, and the result is well known.
The partition function for fluctuation modes of Maxwell soft hair at large $L$ is
\be
Z_F =\prod_{n,\textbf{k}}\left[ \frac{\pi}{\beta^2(\omega_n^2+\textbf{k}^2)}\right]^{1/2}\,,
\ee
where $\omega_n$ and $\textbf{k}$ are Fourier modes along $\tau$ and spatial directions.
The free energy and entropy from this part are read as
\bea
F &=& -T\ln Z_F =-\frac{\zeta(3)}{6\pi} T^3 L^2\,,\nn\\
S_F &=& -\frac{\pd F}{\pd T} = \frac{\zeta(3)}{2\pi}\frac{L^2}{\beta^2}\,.
\eea
Note that $L$ is taken to be large compared to the horizon's scale or inverse temperature here,  mainly for two reasons.
The first one is that, when we are doing asymptotic analysis in the gravity case, we have taken $1/r$ to be small, and the diffeomorphisms we care about are the ones that preserve the asymptotic metric. The extra degrees of freedom are living on the so-called celestial sphere. Here the gauge theory is supposed to have a similar property, and $L$ is taken to be the radius of the celestial sphere.
Second, in order to calculate the partition function of $\phi$, we put those modes in a (2+1)-dimensional box with finite temperature $1/\beta$. The background topology $\mathbb{S}^1\times\mathbb{S}^2$ is taken care of by periodic boundary conditions. Then this is the standard thermal field theory for a scalar field,
\be
\ln Z=-\sum_k [\frac{1}{2} \beta k + \ln (1-e^{-\beta k})]\,,
\ee
with $k=\sqrt{\textbf{k}^2}$.
In order to change the sum into an integral, we need to take a large volume limit. $L\gg\beta$ also justifies that we put the scalar field in a large box.

There are also constant modes or topological modes contribution to the entropy.
The scalar field $\phi$ is compact; thus, the path integral over the zero modes gives out a partition function proportional to the perimeter of the circle up to a normalisation factor.
The corresponding entropy that is proportional to the logarithm of the partition function may dominate when the power-law contribution from thermal modes becomes less important at low temperature.
Also, because we are working on a Euclidean background with compact time direction, the fundamental group element of $S^1$ can be added into the path integral as winding modes.
Those topological modes give out theta function contribution in the partition function.
Again, both contributions from zero mode and topological modes can become important contributions in the low-temperature limit. They may contribute to the logarithm correction of black hole entropy in the near extremal black hole case.
Nevertheless, those contributions are not essential for our discussion here.
The size of the Maxwell soft hair phase space is more or less proportional to
\be
k\propto e^{S_F}=\exp\left(\frac{\zeta(3)}{2\pi}\frac{L^2}{\beta^2}\right)\,.
\ee
Note that $k$ changes as the temperature of the spacetime varies. As the temperature of the spacetime becomes higher and higher, $k$ can be very large.
In other words, the phase space of soft hair increases as more and more Hawking radiation happens.
The variation of $k$ with temperature is the crucial ingredient for our arguments to understand the black hole information paradox, which we will see in the next section.

We do not have an excellent angle to argue what $L$ is in this case.
According to asymptotic analysis, $L$ should be a large distance cutoff, and the boundary should be regarded as the celestial sphere.
Now the transition function prescription in gauge theory does not really care about the falloff conditions, and what we need is just a boundary between two regions. 
In this sense, the boundary can be moved to a finite distance away from the horizon. 
This is the situation that we did not go into too many details and restricted ourselves to large $L$. But it is interesting to consider more about this case.
The scale of $L$ is essential in determining the phase space of soft hair, which will be elaborated on in \cite{Cheng2021}.
The situation is not like the extremal black hole case where we have a natural length scale to characterise the boundary between AdS$_2\times S^2$ throat and asymptotic region.
In principle, $L$ should be determined by length scales $G_N$ and $M$, or even the history of the evaporation, which means $L$ might vary with time.
The point we want to make here is, at the end of the evaporation, where the temperature for the tiny black hole is very high, $k$ should always be large enough to play a significant role.

\section{Page curve from soft hair}
\label{pagecurve}

Having an effective action of Maxwell soft hair at hand, we estimated the size of phase space in the previous section.
In this section, we address the problem of how soft hair of the black hole gives rise to one version of ``Page curve", which might not be the same as the one originally proposed by Page \cite{Page1993}, but certainly does not violate the bound for von Neumann entropy during evaporation.
Note that here the Page curve is gotten from a more microscopic perspective.
The basic idea is that we should at least allow for two kinds of processes on the black hole background, namely, Hawking radiation and measurement.
The measurement is done by a global infalling observer across the boundary.
The competition between those two processes, the rate of which is proportional to the size of phase space, will eventually give out the Page curve.

\subsection{Two types of processes}

In general, we should consider two types of physical processes on a black hole background.
The Hawking radiation process is the well-understood one, which increases black hole entropy by creating the entanglement between the Hawking radiation and interior partner.
There should also be measurement processes, which can be done by U(1) dressed infalling observers.
As argued by PV, the measurement of the soft hair $f$ or $\phi$ enables the reconstruction of Hawking partners, and they also demonstrated the possible quantum information protocol of the reconstruction using measurements of soft hair.
There are still subtleties related to the exchange between code subspace and soft hair subspace, which is not completely clear.
However, naively, one can always say that the comparison of $f$ (or $\phi$ in the Maxwell case) between two regions projects the black hole onto a given soft hair configuration, a lower entropy state.
The entropy of the black hole is reduced once the outside observer in figure \ref{softf} knows more information about what is $f$ or $\phi$.
The soft hair space of the black hole is projected onto a lower entropy state by repeated measurements.
From now on, we will say that the measurement can be done by any infalling dressed particle that crosses the intermediate region, and such measurements reduce the entropy of the black hole.

\begin{figure}
  \centering
  \includegraphics[width=5cm]{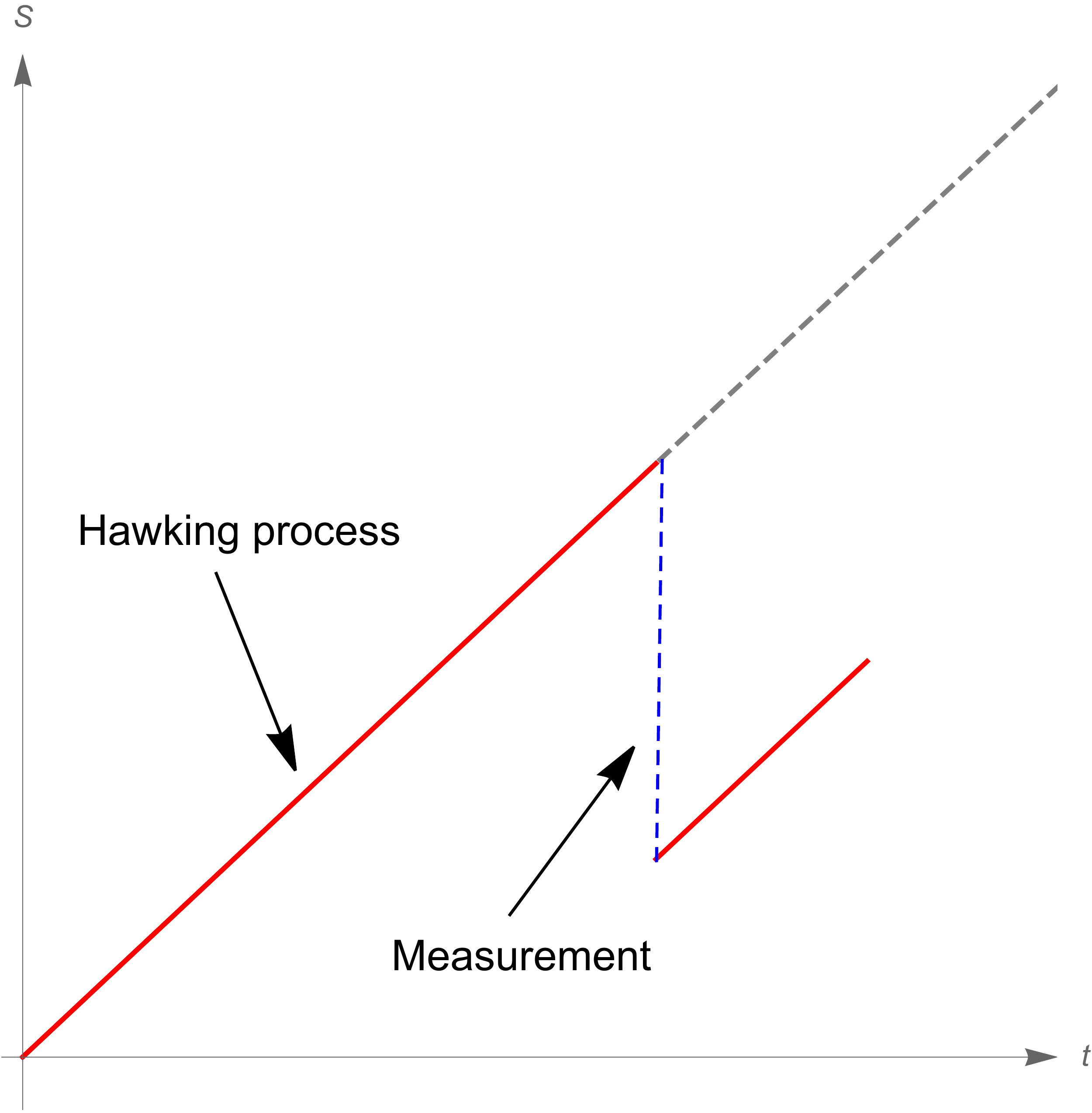}\\
  \caption{The entropy of black hole is increased by Hawking radiation and decreased by measurements that project the black hole onto states with given soft hair configurations.}\label{HvsM}
\end{figure}

The entropy of the black hole is depicted in figure \ref{HvsM}.
As usual, we always assume we are starting with a pure state black hole.
The entropy of the black hole is increased by Hawking radiations, as shown by the red lines in the figure.
The more entanglement between the black hole and Hawking radiation is created, the more entropy the black hole has.
This is precisely Hawking's paradox, which confuses people by saying that the process seems to continue forever, and the entanglement entropy can be even larger than the black hole thermodynamics entropy.
The measurements that reduce the entropy of the black hole are shown by the blue dashed line, which might be invisible to Hawking's argument, but will be the most important new ingredient for the black hole information paradox.
The argument in this section is still very qualitative.
A natural question is how much do the measurements reduce black hole entropy, which requires a competition between the rate of Hawking radiation and soft hair measurement.

\subsection{Fermi's golden rule}

A more accurate version of counting the rate of particle scattering should rely on black hole S matrix \cite{Dray_1985, Hooft1996, Hooft2016}.
There is some recent progress on black hole S matrix related to soft particle dressing \cite{Himwich2020, Gaddam2020, Betzios2020}.
However, for our purpose here, it is already enough to use a very modest version for two-level systems, i.e., ``Fermi's golden rule" \cite{1927a, orear1950nuclear}.
The Fermi's Golden rule means that we can approximately treat Hawking radiation and infalling particles across the boundary at a given energy as a two-level system.
By virtue of Fermi's golden rule, the rate of transition $w$ can be written as
\be
w=\langle\mathcal{T}\rangle^2 \rho(E)\,,\label{rates}
\ee
with $\mathcal{T}$ as the scattering matrix of the process, and $\rho(E)$ as the density of state of energy $E$. We can use the formula (\ref{rates}) to calculate the rate of Hawking radiation and the measurement process. Then the whole problem is reduced to a density of state counting, assuming that $\langle \mathcal{T}\rangle$ is more or less of the same order.

The intuitive way of understanding Fermi's golden rule is following.
For the black hole with mass $M$ emitting a Hawking quantum with energy $\delta M$ and then going to a lower energy state, we can naturally portray it as a two-level system.
Of course, the rate for this process depends on the density of state of the given black hole state.
The more states we have at the given energy of the black hole, the bigger chance Hawking radiation can happen.
Similarly, for the infalling particles waiting to cross the boundary, one should think this process is a domain-wall-crossing process, which can also be portrayed as a two-level system.
More degeneracy at a given temperature means one has more choices from to choose, and it is easier to make the deal.

Let us first look at the Hawking radiation process.
The density of state can be written as an exponent of the Bekenstein-Hawking entropy $e^{S_{BH}}$.
Then the rate of a black hole with mass $M$ emitting particles at the horizon can be written as \cite{Hooft1985}
\be
w_H=\langle \mathcal{T}_H\rangle^2 ~e^{S_{BH}(M)}\,.
\ee
where $w_H$ is the rate of our first physical process Hawking radiation.
For the second process, the phase space of degeneracy at a given temperature was calculated in the previous section, where we denoted it as
\be
k\propto \exp\left(\frac{\zeta(3)}{2\pi}\frac{L^2}{\beta^2}\right)\,.
\ee
Then the rate of measurement can be written as
\be
w_M=\langle \mathcal{T}_M\rangle^2 \exp\left(\frac{\zeta(3)}{2\pi}\frac{L^2}{\beta^2}\right)\,.
\ee
We need more details of the interacting Hamiltonian to calculate the scattering matrix.
For now, let us assume the overall scattering amplitude of those two processes is comparable with each other.
Then one can compare the rate of those processes just based on the size of phase space.

Those two densities of states are both exponents of entropy.
At the early stage of Hawking radiation where we have a low-temperature black hole, the rate of measurements can be minimal.
In this period, Hawking radiation plays a dominant role.
whereas as the temperature gets higher and higher, it is possible to have
\be
k> e^{S_{BH}}\,.
\ee
Then the measurement process, which decreases the entropy of the black hole, plays a significant role. The measurement of the soft hair does not mean the possibility of Hawking radiation being reduced; the entropy of the radiation still goes as Hawking's calculation.
That only means we can have a large amount of ``invisible" processes that may reduce the entropy of the black hole by measuring the soft hair degrees of freedom.

\subsection{Page curve}

In this subsection, we draw the Page curve of black hole entropy by including the physical effects of measurement of soft hair.

Hawking's calculation of entanglement pair creations should not be changed too much and go until the end of the evaporation.
The new ingredient is the measurement process.

\begin{figure}
  \centering
  \includegraphics[width=4.5cm]{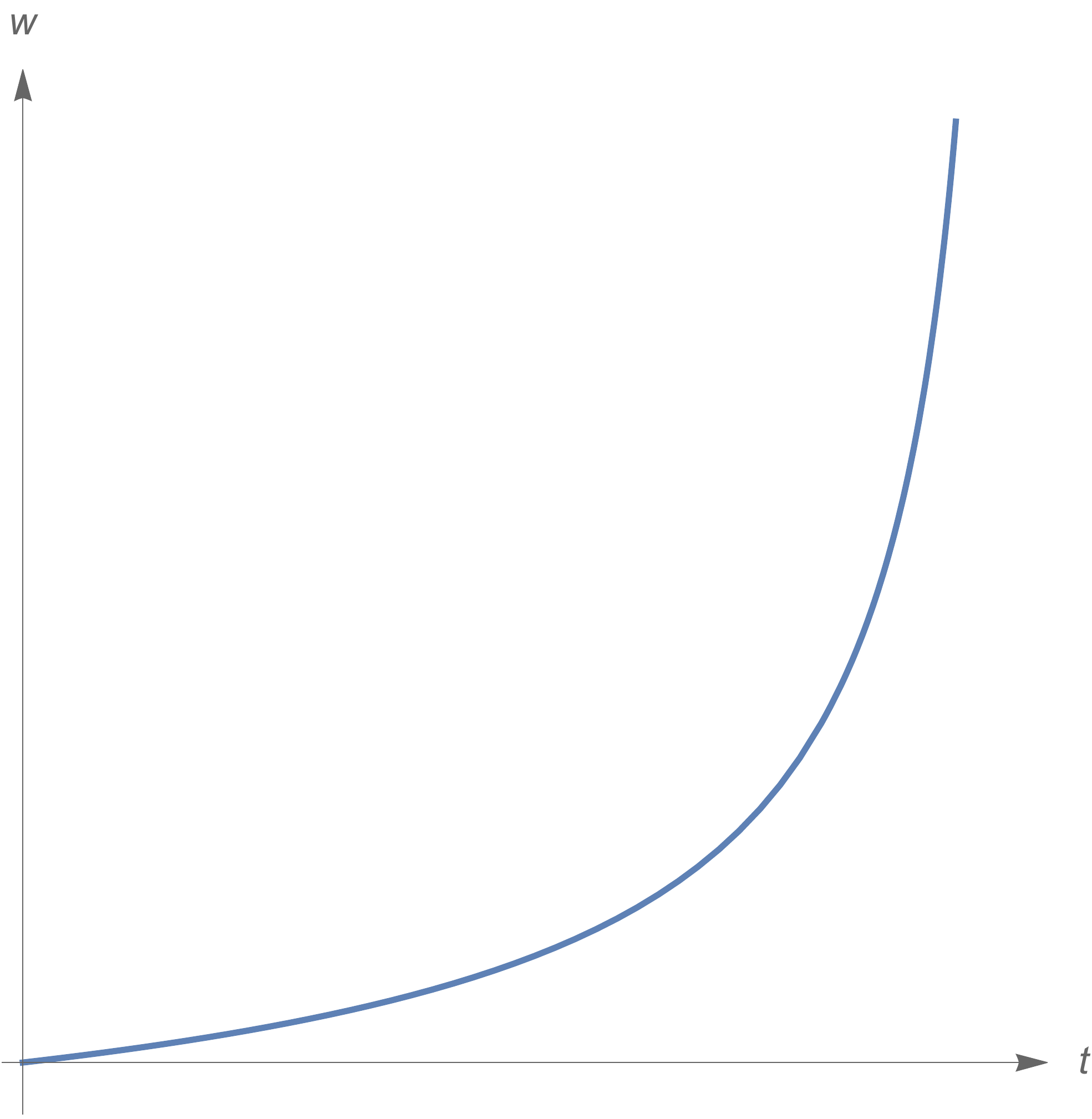}\\
  \caption{A sketch of the rate of measurement $w$ under the assumption of constant $L$ during the evaporation.}\label{rate}
\end{figure}

\begin{figure}
  \centering
  \includegraphics[width=7cm]{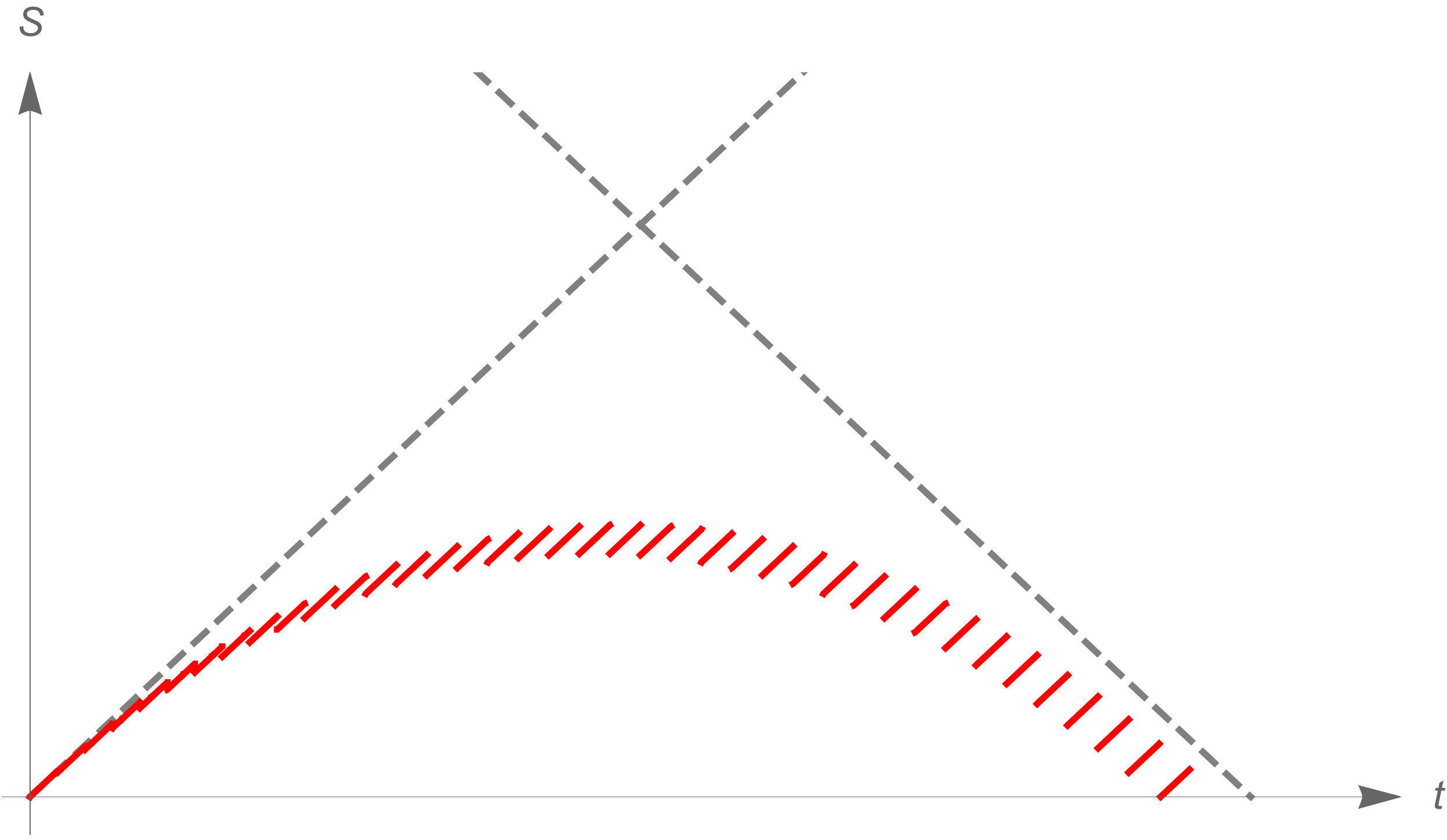}
  \includegraphics[width=7cm]{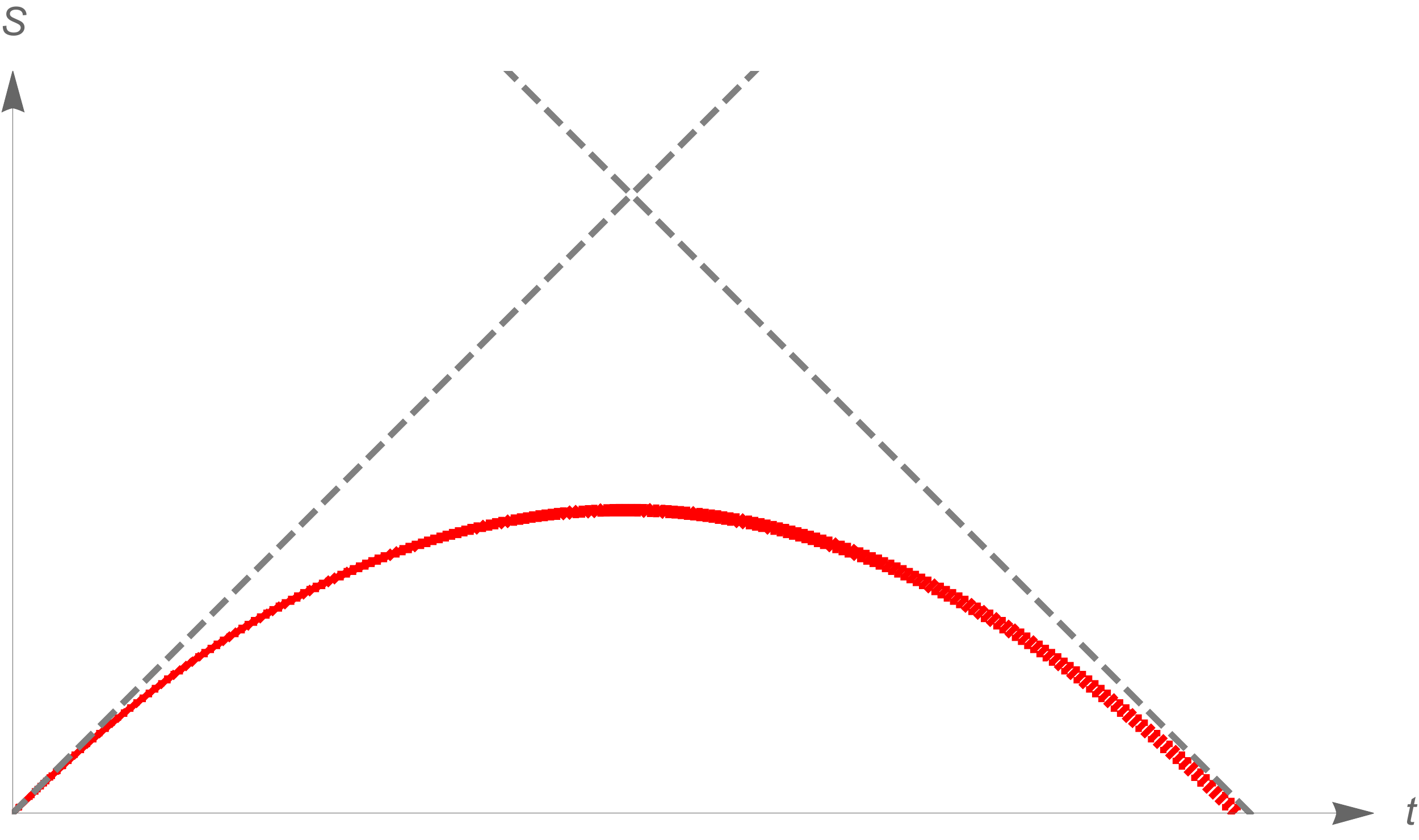}\\
  \caption{Page curve of black hole entropy with the effect of measurements of soft hair being included. The phase space of soft hair is assumed to have a linear growth with time. The first figure is a coarse-grained version, whereas in the second figure the Page curve is ``smoothed" by taking smaller steps in time. }\label{page}
\end{figure}

The central insight we have from previous subsections is that the phase space of this process increases with time.
Without an expression for $L$ at hand, let us assume linear growth of soft hair phase space with time for simplicity
\footnote{
The explicit phase space always depends on the choice of ensemble.
Nevertheless, we can work out the time dependence of density of states by ignoring the time dependence of $L$ and only considering the temperature change because of Hawking radiation.
Then the density of state of soft hair is proportional to exp($T^2(t)$).
If the Hawking radiation $dM/dt=-\alpha/M^2$ with constant $\alpha$, the shape of measurement rate can be depicted in figure \ref{rate}, which means that the modification from the global effects is relatively low at the beginning of the radiation, then becomes greatly enhanced at late times.
}.
Again for simplicity, let us assume the Hawking radiation is emitted at a constant rate, and one infalling particle crossing the intermediate region does not change the energy of the black hole but reduces the entropy of the black hole by one unit.
Now the measurement rate also has a linear growth in time. The Page curve of linear-growth measurement phase space is illustrated in figure \ref{page}.
The red line is continuously growing with time because of Hawking radiation creating entanglement.
The measurements of soft hair decrease the entropy as shown in Fig. \ref{HvsM}; that is the reason for discontinuity in the first panel of figure \ref{page}.
In the second panel of figure \ref{page}, we take smaller steps in numerical approximation, which gives a more ``smooth" curve.

One needs to be aware that those curves are not perfect because of the simplicity assumption we have taken, and if the time dependence of measurement looks like the curve shown Fig. \ref{rate}, the Page curve will look different. The modification of Hawking's result is relatively small at the early time, so the curve more or less follows a straight line. At the late time, the entropy of the black hole gains a more significant modification from this global measurement, then the entropy quickly drops to zero. Then the curve is more or less the Page curve one expected, and those curves are all consistent with the unitary evolution of the black hole.

Note that our derivation of the Page curve has no contradiction with Hawking's calculation.
All the red lines in figure \ref{page} are precisely coming from Hawking's calculation.
We are just saying that there are physical effects invisible to Hawking and missed by Hawking's calculation.
By adding those global effects back to the black hole, the entropy of the black hole can be extensively reduced, thus there is no violation of unitarity during the evaporation.
That means that, to get a convincing Page curve, one might not need a full quantum theory of gravity, but global effects are indeed needed.
The right way to derive a Page curve for radiation, in addition to just saying the whole system is pure, needs further study.

The Page curve we have shown in the second panel of figure \ref{page} looks like a smooth curve, rather than the usually expected phase transition at Page time $t_{\text{page}}$.
Actually, this curve is not a smooth but a wavy curve, which means that if one enlarges each point there is the same pattern as shown in figure \ref{HvsM}.
The wavy curve is because there are many phase transitions at each measurement, as demonstrated in figure \ref{HvsM}.
It is worth mentioning that this enormous amount of phase transitions, but is not strange but getting popular recently because of the island prescription.
It has believed that in the calculation of the entropy of black hole or radiation using Euclidean path integral, there are also a large number of other saddles between the fully connected and disconnected geometries \cite{Penington2019a}.
So there should not be only one phase transition, but many relatively small phase transitions between different saddles.
Here we show the similar effects in our Page curve shown in Fig. \ref{page}.

\section{Conclusion and discussion}
\label{conclusion}

Following the same strategy as \cite{Pasterski2020}, we treated the Maxwell soft hair degrees of freedom as a transition function that relates the Maxwell gauge fields in the asymptotic region and near-horizon region.
We introduced U(1) gauge field $A_{\mu}$ living in the near-horizon region and $\bar A_{\mu}$ in the asymptotic flat region, which are related by a gauge transformation $\bar A = A+\dd\phi$ on the boundary between those two regions.
The transition function $\phi$, i.e. would-be gauge degrees of freedom, is regarded as the Goldstone modes that characterise the Maxwell soft hair. The Maxwell edge modes and soft hair are naturally connected.

The advantage of the U(1) gauge case is that an effective action description of those soft hair degrees of freedom can be easily obtained. By separating the bulk fluctuation modes that vanish on the boundary and boundary soft hair modes that are on shell in bulk, we get the effective action for soft hair which is just the one for massless scalar field
\be
S[\phi]=-\frac{1}{2 f(L)}\int_{\pmM} d^3x \sqrt{-h}~ \pd^a\phi\pd_a\phi\,.
\ee
Thus one can use the Euclidean path integral in terms of the effective action to study the statistical properties of the Maxwell soft hair.
The entropy gotten from the path integral is roughly $S_{\text{soft}} \propto L^2/\beta^2$.
The phase space of soft hair increases as the temperature of the black hole becomes higher and higher.

There should be two basic processes on the black hole background: Hawking radiation and measurement of soft hair.
The measurement of soft hair can be done by the infalling observer who measures the U(1) phase difference between two sides.
Those global effects are invisible to local observers.
It was shown in the paper by PV that the measurement implements a code subspace projection that enables the reconstruction of interior operators.
Fermi's golden rule states that, for two-level systems, the rate of changing energy level is proportional to the density of states at each level.
At a given energy, we can regard the Hawking radiation and boundary-crossing process as two-level systems.
The competition of the size of phase spaces would finally give rise to the Page curve.
At the early stage, the Hawking radiations have a large phase space, and thus they dominate.
While at the late stage, the boundary-crossing measurements becoming important and thus reduce the entropy of the black hole.
One version of the Page curve is shown in figure \ref{page}.

Our results suggest that soft hair as a global effect might play an essential role in the unitary evolution of black holes and be powerful enough to give the Page curve of black hole evolution.
Is Hawking's calculation of radiation valid until the last breath of the black hole? The answer might be yes, and even so, there is no contradiction with the Page curve.

Understanding the exact location of the boundary between the near-horizon region and asymptotic region is essential in determining the phase space of soft hair.
The critical question for the brave infalling observer jumping into the black hole is when will the phase difference between the asymptotic region and near-horizon region start to be felt.
We tend to believe $L$ is an infrared cutoff at a large distance because of falloff analysis of asymptotic symmetry \cite{Strominger2017}.
One might argue why crossing a boundary at a large distance can do anything to the black hole. This question might be answered by ER=EPR or emphasising that those are global effects of the black hole.

All the arguments above are based on the assumption that measurements of the soft hair enable the reconstruction of Hawking partners.
Although we have seen some evidence of such reconstruction \cite{Pasterski2020, Yoshida2019}, there is no clear argument about how the exchange between the code subspace and soft hair subspace could happen.
In other words, if the information of the Hawking partner is stored in the soft hair degrees of freedom, which is invisible for local Hawking calculation, what is the physical process of information exchange between Hawking radiation and soft hair?
Does this happen when the partner crosses the horizon or when the Hawking radiation crosses the boundary $r=L$?
Such questions are always the critical ingredients of understanding the problem, which might be very hard to answer within the current framework.

It would be very interesting to understand the connection between the soft hair story and island prescription.
One clue of the relation is how to understand the replica wormholes arising from Euclidean path integral derivation of island rule \cite{Penington2019a, Almheiri2019b}.
Should those replica wormholes be understood as some kind of domain-wall-crossing process, it might be easier to build the connection.
They indeed share a lot of similarity in terms of phase space counting to decide which process dominates, reconstruction of the interior partner of Hawking radiation, and so on.
Another interesting question to understand is that, if the soft hair provides extra parameters labelling superselection sectors, does swampland have something to say about those ``soft symmetries"?

\section*{Acknowledgements}
We would like to thank Ankit Aggarwal, Yu-Sen An, Jan de Boer, Nava Gaddam, Bin Guo, Diego Hofman, Shuan-Ming Ruan, and Zhenbin Yang for inspirational discussions. P. C. is financially supported by China Scholarship Council (CSC).


\begin{thebibliography}{100}

\bibitem{Hawking1976}
S.W.~Hawking, \emph{Breakdown of predictability in gravitational collapse},
  \href{https://doi.org/10.1103/physrevd.14.2460}{\emph{Phys Rev D} {\bfseries
  14} (1976) 2460}.

\bibitem{HAWKING1974}
S.W.~Hawking, \emph{Black hole explosions?},
  \href{https://doi.org/10.1038/248030a0}{\emph{Nature} {\bfseries 248} (1974)
  30}.

\bibitem{Hawking1975}
S.W.~Hawking, \emph{Particle creation by black holes},
  \href{https://doi.org/10.1007/bf02345020}{\emph{Communications In
  Mathematical Physics} {\bfseries 43} (1975) 199}.

\bibitem{Penington2019}
G.~Penington, \emph{Entanglement wedge reconstruction and the information
  paradox},  \href{https://arxiv.org/abs/1905.08255v2}{{\ttfamily
  1905.08255v2}}.

\bibitem{Almheiri2019}
A.~Almheiri, N.~Engelhardt, D.~Marolf and H.~Maxfield, \emph{The entropy of
  bulk quantum fields and the entanglement wedge of an evaporating black hole},
  \href{https://doi.org/10.1007/JHEP12(2019)063}{\emph{J High Energy Phys}
  {\bfseries 2019} (2019) }
  [\href{https://arxiv.org/abs/1905.08762v3}{{\ttfamily 1905.08762v3}}].

\bibitem{Almheiri2019a}
A.~Almheiri, R.~Mahajan, J.~Maldacena and Y.~Zhao, \emph{The Page curve of
  Hawking radiation from semiclassical geometry},
  \href{https://doi.org/10.1007/JHEP03(2020)149}{\emph{J High Energy Phys}
  {\bfseries 2020} (2020) }
  [\href{https://arxiv.org/abs/1908.10996v2}{{\ttfamily 1908.10996v2}}].

\bibitem{Almheiri2019c}
A.~Almheiri, R.~Mahajan and J.~Maldacena, \emph{Islands outside the horizon},
  \href{https://arxiv.org/abs/1910.11077v3}{{\ttfamily 1910.11077v3}}.

\bibitem{Penington2019a}
G.~Penington, S.H.~Shenker, D.~Stanford and Z.~Yang, \emph{Replica wormholes
  and the black hole interior},
  \href{https://arxiv.org/abs/1911.11977v2}{{\ttfamily 1911.11977v2}}.

\bibitem{Almheiri2019b}
A.~Almheiri, T.~Hartman, J.~Maldacena, E.~Shaghoulian and A.~Tajdini,
  \emph{Replica wormholes and the entropy of Hawking radiation},
  \href{https://doi.org/10.1007/JHEP05(2020)013}{\emph{J High Energy Phys}
  {\bfseries 2020} (2020) }
  [\href{https://arxiv.org/abs/1911.12333v2}{{\ttfamily 1911.12333v2}}].

\bibitem{Almheiri2020}
A.~Almheiri, T.~Hartman, J.~Maldacena, E.~Shaghoulian and A.~Tajdini, \emph{The
  entropy of Hawking radiation},
  \href{https://arxiv.org/abs/2006.06872v1}{{\ttfamily 2006.06872v1}}.

\bibitem{Almheiri2012}
A.~Almheiri, D.~Marolf, J.~Polchinski and J.~Sully, \emph{Black holes:
  Complementarity or firewalls?},
  \href{https://doi.org/10.1007/JHEP02(2013)062}{\emph{J High Energy Phys}
  {\bfseries 2013} (2013) } [\href{https://arxiv.org/abs/1207.3123}{{\ttfamily
  1207.3123}}].

\bibitem{Israel1967}
W.~Israel, \emph{Event horizons in static vacuum space-times},
  \href{https://doi.org/10.1103/physrev.164.1776}{\emph{Phys Rev} {\bfseries
  164} (1967) 1776}.

\bibitem{Israel1968}
W.~Israel, \emph{Event horizons in static electrovac space-times},
  \href{https://doi.org/10.1007/bf01645859}{\emph{Communications in
  Mathematical Physics} {\bfseries 8} (1968) 245}.

\bibitem{Papadodimas2012}
K.~Papadodimas and S.~Raju, \emph{An infalling observer in AdS/CFT},
  \href{https://doi.org/10.1007/JHEP10(2013)212}{\emph{J High Energy Phys}
  {\bfseries 2013} (2013) } [\href{https://arxiv.org/abs/1211.6767}{{\ttfamily
  1211.6767}}].

\bibitem{Verlinde2012}
E.~Verlinde and H.~Verlinde, \emph{Black hole entanglement and quantum error
  correction}, \href{https://doi.org/10.1007/JHEP10(2013)107}{\emph{J High
  Energy Phys} {\bfseries 2013} (2013) }
  [\href{https://arxiv.org/abs/1211.6913}{{\ttfamily 1211.6913}}].

\bibitem{Pasterski2020}
S.~Pasterski and H.~Verlinde, \emph{HPS meets AMPS: How soft hair dissolves the
  firewall},  \href{https://arxiv.org/abs/2012.03850}{{\ttfamily 2012.03850}}.

\bibitem{Donnelly2014}
W.~Donnelly and A.C.~Wall, \emph{Entanglement entropy of electromagnetic edge
  modes}, \href{https://doi.org/10.1103/PhysRevLett.114.111603}{\emph{Phys Rev
  Lett} {\bfseries 114} (2015) }
  [\href{https://arxiv.org/abs/1412.1895v2}{{\ttfamily 1412.1895v2}}].

\bibitem{Donnelly2015}
W.~Donnelly and A.C.~Wall, \emph{Geometric entropy and edge modes of the
  electromagnetic field},
  \href{https://doi.org/10.1103/PhysRevD.94.104053}{\emph{Phys Rev D}
  {\bfseries 94} (2016) } [\href{https://arxiv.org/abs/1506.05792v1}{{\ttfamily
  1506.05792v1}}].

\bibitem{Donnelly2016}
W.~Donnelly and L.~Freidel, \emph{Local subsystems in gauge theory and
  gravity}, \href{https://doi.org/10.1007/JHEP09(2016)102}{\emph{J High Energy
  Phys} {\bfseries 2016} (2016) }
  [\href{https://arxiv.org/abs/1601.04744v2}{{\ttfamily 1601.04744v2}}].

\bibitem{Blommaert2018}
A.~Blommaert, T.G.~Mertens, H.~Verschelde and V.I.~Zakharov, \emph{Edge state
  quantization: Vector fields in rindler},
  \href{https://doi.org/10.1007/JHEP08(2018)196}{\emph{J High Energy Phys}
  {\bfseries 2018} (2018) }
  [\href{https://arxiv.org/abs/1801.09910v2}{{\ttfamily 1801.09910v2}}].

\bibitem{Blommaert2018a}
A.~Blommaert, T.G.~Mertens and H.~Verschelde, \emph{Edge dynamics from the path
  integral: Maxwell and Yang-Mills},
  \href{https://doi.org/10.1007/JHEP11(2018)080}{\emph{J High Energy Phys}
  {\bfseries 2018} (2018) }
  [\href{https://arxiv.org/abs/1804.07585v2}{{\ttfamily 1804.07585v2}}].

\bibitem{Hawking2015}
S.W.~Hawking, \emph{The information paradox for black holes},
  \href{https://arxiv.org/abs/1509.01147v1}{{\ttfamily 1509.01147v1}}.

\bibitem{Hawking2016a}
S.W.~Hawking, M.J.~Perry and A.~Strominger, \emph{Superrotation charge and
  supertranslation hair on black holes},
  \href{https://doi.org/10.1007/JHEP05(2017)161}{\emph{J High Energy Phys}
  {\bfseries 2017} (2017) }
  [\href{https://arxiv.org/abs/1611.09175v2}{{\ttfamily 1611.09175v2}}].

\bibitem{Strominger2017}
A.~Strominger, \emph{Lectures on the infrared structure of gravity and gauge
  theory},  \href{https://arxiv.org/abs/1703.05448v2}{{\ttfamily
  1703.05448v2}}.

\bibitem{Strominger2017a}
A.~Strominger, \emph{Black hole information revisited},
  \href{https://arxiv.org/abs/1706.07143v1}{{\ttfamily 1706.07143v1}}.

\bibitem{Mirbabayi2016}
M.~Mirbabayi and M.~Porrati, \emph{Shaving off black hole soft hair},
  \href{https://doi.org/10.1103/PhysRevLett.117.211301}{\emph{Phys Rev Lett}
  {\bfseries 117} (2016) } [\href{https://arxiv.org/abs/1607.03120}{{\ttfamily
  1607.03120}}].

\bibitem{Bousso2017}
R.~Bousso and M.~Porrati, \emph{Soft hair as a soft wig},
  \href{https://doi.org/10.1088/1361-6382/aa8be2}{\emph{Classical Quantum
  Gravity} {\bfseries 34} (2017) 204001}
  [\href{https://arxiv.org/abs/1706.00436}{{\ttfamily 1706.00436}}].

\bibitem{Haco2018}
S.~Haco, S.W.~Hawking, M.J.~Perry and A.~Strominger, \emph{Black hole entropy
  and soft hair}, \href{https://doi.org/10.1007/JHEP12(2018)098}{\emph{J High
  Energy Phys} {\bfseries 2018} (2018) }
  [\href{https://arxiv.org/abs/1810.01847v4}{{\ttfamily 1810.01847v4}}].

\bibitem{Haco2019}
S.~Haco, M.J.~Perry and A.~Strominger, \emph{Kerr-Newman black hole entropy and
  soft hair},  \href{https://arxiv.org/abs/1902.02247v1}{{\ttfamily
  1902.02247v1}}.

\bibitem{Donnay2018}
L.~Donnay, G.~Giribet, H.A.~Gonzalez and A.~Puhm, \emph{Black hole memory
  effect}, \href{https://doi.org/10.1103/PhysRevD.98.124016}{\emph{Phys Rev D}
  {\bfseries 98} (2018) } [\href{https://arxiv.org/abs/1809.07266v1}{{\ttfamily
  1809.07266v1}}].

\bibitem{Nomura2018}
Y.~Nomura, \emph{Reanalyzing an evaporating black hole},
  \href{https://doi.org/10.1103/PhysRevD.99.086004}{\emph{Phys Rev D}
  {\bfseries 99} (2019) } [\href{https://arxiv.org/abs/1810.09453}{{\ttfamily
  1810.09453}}].

\bibitem{Nomura2019}
Y.~Nomura, \emph{The interior of a unitarily evaporating black hole},
  \href{https://doi.org/10.1103/PhysRevD.102.026001}{\emph{Phys Rev D}
  {\bfseries 102} (2020) } [\href{https://arxiv.org/abs/1911.13120}{{\ttfamily
  1911.13120}}].

\bibitem{Betzios2020}
P.~Betzios, N.~Gaddam and O.~Papadoulaki, \emph{Black hole S-matrix for a
  scalar field},  \href{https://arxiv.org/abs/2012.09834}{{\ttfamily
  2012.09834}}.

\bibitem{Gaddam2020}
N.~Gaddam and N.~Groenenboom, \emph{Soft graviton exchange and the information
  paradox},  \href{https://arxiv.org/abs/2012.02355}{{\ttfamily 2012.02355}}.

\bibitem{Himwich2020}
E.~Himwich, S.A.~Narayanan, M.~Pate, N.~Paul and A.~Strominger, \emph{The soft
  $\mathcal{S}$-matrix in gravity},
  \href{https://doi.org/10.1007/JHEP09(2020)129}{\emph{J High Energy Phys}
  {\bfseries 2020} (2020) }
  [\href{https://arxiv.org/abs/2005.13433v1}{{\ttfamily 2005.13433v1}}].

\bibitem{Verlinde2020}
H.~Verlinde, \emph{ER = EPR revisited: On the entropy of an Einstein-Rosen
  bridge},  \href{https://arxiv.org/abs/2003.13117}{{\ttfamily 2003.13117}}.

\bibitem{Marolf2020}
D.~Marolf and H.~Maxfield, \emph{Observations of Hawking radiation: the Page
  curve and baby universes},
  \href{https://arxiv.org/abs/2010.06602}{{\ttfamily 2010.06602}}.

\bibitem{Marolf2020a}
D.~Marolf and H.~Maxfield, \emph{Transcending the ensemble: baby universes,
  spacetime wormholes, and the order and disorder of black hole information},
  \href{https://doi.org/10.1007/jhep08(2020)044}{\emph{J High Energy Phys}
  {\bfseries 2020} (2020) } [\href{https://arxiv.org/abs/2002.08950}{{\ttfamily
  2002.08950}}].

\bibitem{Saad2019}
P.~Saad, S.H.~Shenker and D.~Stanford, \emph{JT gravity as a matrix integral},
  \href{https://arxiv.org/abs/1903.11115}{{\ttfamily 1903.11115}}.

\bibitem{Bousso2020}
R.~Bousso and E.~Wildenhain, \emph{Gravity/Ensemble duality},
  \href{https://arxiv.org/abs/2006.16289v2}{{\ttfamily 2006.16289v2}}.

\bibitem{Harlow2018}
D.~Harlow and H.~Ooguri, \emph{Symmetries in quantum field theory and quantum
  gravity},  \href{https://arxiv.org/abs/1810.05338}{{\ttfamily 1810.05338}}.

\bibitem{Harlow2018a}
D.~Harlow and H.~Ooguri, \emph{Constraints on symmetry from holography},
  \href{https://arxiv.org/abs/1810.05337}{{\ttfamily 1810.05337}}.

\bibitem{McNamara2020}
J.~McNamara and C.~Vafa, \emph{Baby universes, holography, and the swampland},
  \href{https://arxiv.org/abs/2004.06738}{{\ttfamily 2004.06738}}.

\bibitem{Harlow2020}
D.~Harlow and E.~Shaghoulian, \emph{Global symmetry, euclidean gravity, and the
  black hole information problem},
  \href{https://arxiv.org/abs/2010.10539}{{\ttfamily 2010.10539}}.

\bibitem{Fichet2019}
S.~Fichet and P.~Saraswat, \emph{Approximate symmetries and gravity},
  \href{https://arxiv.org/abs/1909.02002}{{\ttfamily 1909.02002}}.

\bibitem{Cheng2021}
J.~de~Boer, P.~Cheng and D.~Hofman, \emph{To appear}, .

\bibitem{Page1993}
D.N.~Page, \emph{Information in black hole radiation},
  \href{https://doi.org/10.1103/PhysRevLett.71.3743}{\emph{Phys Rev Lett}
  {\bfseries 71} (1993) 3743}
  [\href{https://arxiv.org/abs/hep-th/9306083v2}{{\ttfamily
  hep-th/9306083v2}}].

\bibitem{Dray_1985}
T.~Dray and G.~'t~Hooft, \emph{The gravitational shock wave of a massless
  particle}, \href{https://doi.org/10.1016/0550-3213(85)90525-5}{\emph{Nucl
  Phys B} {\bfseries 253} (1985) 173}.

\bibitem{Hooft1996}
G.~'t~Hooft, \emph{The scattering matrix approach for the quantum black hole,
  an overview},
  \href{https://doi.org/10.1142/S0217751X96002145}{\emph{International Journal
  of Modern Physics A} {\bfseries 11} (1996) 4623}
  [\href{https://arxiv.org/abs/gr-qc/9607022v1}{{\ttfamily gr-qc/9607022v1}}].

\bibitem{Hooft2016}
G.~'t~Hooft, \emph{Black hole unitarity and antipodal entanglement},
  \href{https://doi.org/10.1007/s10701-016-0014-y}{\emph{Found Phys} {\bfseries
  46} (2016) 1185} [\href{https://arxiv.org/abs/1601.03447v4}{{\ttfamily
  1601.03447v4}}].

\bibitem{1927a}
P.A.~Dirac, \emph{The quantum theory of the emission and absorption of
  radiation}, \href{https://doi.org/10.1098/rspa.1927.0039}{\emph{Proceedings
  of the Royal Society of London. Series A, Containing Papers of a Mathematical
  and Physical Character} {\bfseries 114} (1927) 243}.

\bibitem{orear1950nuclear}
E.~Fermi, \emph{Nuclear Physics: A Course Given by Enrico Fermi at the
  University of Chicago}, Midway reprint, University of Chicago Press (1950).

\bibitem{Hooft1985}
G.~‘t Hooft, \emph{On the quantum structure of a black hole},
  \href{https://doi.org/10.1016/0550-3213(85)90418-3}{\emph{Nucl Phys B}
  {\bfseries 256} (1985) 727}.

\bibitem{Yoshida2019}
B.~Yoshida, \emph{Observer-dependent black hole interior from operator
  collision},  \href{https://arxiv.org/abs/1910.11346}{{\ttfamily 1910.11346}}.

\end{thebibliography}

\providecommand{\href}[2]{#2}\begingroup\raggedright\endgroup

\end{document}